\documentclass[conference]{IEEEtran}

\usepackage{url}

\usepackage{tikz}
\usepackage{amsmath, xspace}

\makeatletter
\newenvironment{customitemize}{%
  \begin{list}{$\bullet$}{%
    \setlength{\itemsep}{3pt}%
    \setlength{\parsep}{0pt}%
    \setlength{\topsep}{0pt}%
    \setlength{\partopsep}{0pt}%
    \setlength{\leftmargin}{2.5em}
    \setlength{\labelwidth}{1em}%
    \setlength{\labelsep}{0.5em}%
  }%
}{%
  \end{list}%
}
\makeatother

\newcounter{customsection}

\usepackage{ulem}
\usepackage{epsfig,amsmath,amsfonts,multirow,makecell,caption,soul,csquotes,color,wrapfig,subcaption,mathtools,bm,spverbatim,booktabs,xcolor,amsthm,tcolorbox,wrapfig, bbm, cite}
\usepackage{diagbox}
\usepackage{xcolor,colortbl}
\usepackage{tikz}
\usepackage[hidelinks]{hyperref}

\usepackage{amssymb}
\usepackage{pifont}
\newcommand{\cmark}{\ding{51}}%
\newcommand{\xmark}{\ding{55}}%
\newcounter{countobs}
\newcounter{refcounter}

\newcommand{\newparts}[1]{{#1}}
\newcommand{\cut}[1]{}

\newcommand\bluesout{\bgroup\markoverwith{\textcolor{blue}{\rule[0.5ex]{2pt}{0.4pt}}}\ULon}

\newcommand{\metricHA}{$f_\mathrm{HA}$\xspace} 
\newcommand{\metricAA}{$f_\mathrm{AA}$\xspace} 

\usepackage{mdframed}

\newtcolorbox{mtbox}[1]{
  left=0.5mm,
  right=0.5mm,
  top=0.5mm,
  bottom=0.5mm,
  colframe=black, 
  colback=white, 
  boxrule=0.5pt,
  title={#1},
  fonttitle=\bfseries,
  coltitle=black,
  attach title to upper={},
  width=\linewidth,
  sharp corners
}

\usepackage{mdframed}
\mdfdefinestyle{rebuttalstyle}{innerleftmargin=3pt, innerrightmargin=3pt, innertopmargin=3pt, innerbottommargin=3pt}

\newcounter{link}[section]

\newcounter{response}[section]

\newcounter{revision}[section]

\ifCLASSINFOpdf
\else
\fi
\begin{document}
%
\title{Revisiting Physical-World Adversarial Attack on Traffic Sign Recognition: A Commercial Systems Perspective}

\author{{\rm Ningfei Wang}  \quad {\rm Shaoyuan Xie} \quad {\rm Takami Sato} \quad {\rm Yunpeng Luo} \quad {\rm Kaidi Xu}$^{\dagger}$ \quad {\rm Qi Alfred Chen} \\
University of California, Irvine \quad
$^\dagger$Drexel University\\
\{ningfei.wang, shaoyux, takamis, yunpel3, alfchen\}@uci.edu \quad $^\dagger$kx46@drexel.edu}

\IEEEoverridecommandlockouts
\makeatletter\def\@IEEEpubidpullup{6.5\baselineskip}\makeatother
\IEEEpubid{\parbox{\columnwidth}{
		Network and Distributed System Security (NDSS) Symposium 2025\\
		23 - 28 February 2025, San Diego, CA, USA\\
		ISBN 979-8-9894372-8-3\\
		https://dx.doi.org/10.14722/ndss.2025.23090\\
		www.ndss-symposium.org
}
\hspace{\columnsep}\makebox[\columnwidth]{}}

\maketitle



%

\begin{abstract}

Traffic Sign Recognition (TSR) is crucial for safe and correct driving automation. Recent works revealed a general vulnerability of TSR models to physical-world adversarial attacks, which can be low-cost, highly deployable, and capable of causing severe attack effects such as hiding a critical traffic sign or spoofing a fake one. However, so far existing works generally only considered evaluating the attack effects on academic TSR models, leaving the impacts of such attacks on real-world commercial TSR systems largely unclear. In this paper, we conduct the first large-scale measurement of physical-world adversarial attacks against commercial TSR systems. Our testing results reveal that it is possible for existing attack works from academia to have highly reliable (100\%) attack success against certain commercial TSR system functionality, but such attack capabilities are not generalizable, leading to much lower-than-expected attack success rates overall. We find that one potential major factor is a spatial memorization design that commonly exists in today's commercial TSR systems. We design new attack success metrics that can mathematically model the impacts of such design on the TSR system-level attack success, and use them to revisit existing attacks. Through these efforts, we uncover 7 novel observations, some of which directly challenge the observations or claims in prior works due to the introduction of the new metrics.

\end{abstract}

\section{Introduction}
\label{sec:intro}
In the rapidly evolving landscape of Autonomous Driving (AD) technology, AD vehicles, such as the millions of Tesla cars~\cite{Kane2021Tesla} running on the public road, are becoming an integral part of our daily lives. Compliance with traffic signs is essential for all vehicles, no matter if they are high-autonomy AD vehicles (e.g., those for robo-taxi~\cite{waymotaxi}), semi-autonomous AD vehicles (e.g., those with Tesla Autopilot~\cite{teslafuture}), or conventional human-driven vehicles. Failure to obey these rules can lead to accidents, posing a threat to human life.

Due to the importance of traffic sign detection, a natural question is whether AD vehicles are truly as secure as we hope. To answer this critical question, recent research in security analysis of Traffic Sign Recognition (TSR) systems has highlighted vulnerabilities to a wide range of physical adversarial attacks that can significantly impair the traffic sign detection accuracy~\cite{lu2017adversarial, zhao2019seeing, Wang_2023_ICCV, jia2022fooling, nassi2020phantom, eykholt2018physical, chen2018shapeshifter, sato2023intriguing, lovisotto2021slap, zhong2022shadows}. Among them, the most representative and also the most widely-exploited attack vectors are \textit{physical patches or posters}~\cite{lu2017adversarial, Wang_2023_ICCV, zhao2019seeing, eykholt2018physical, jia2022fooling, chen2018shapeshifter, sato2023intriguing}, which are low-cost, highly deployable, and demonstrated capable of causing various highly severe attack effects. For instance, they can make critical legitimate traffic signs undetectable, or \textit{hiding attacks}, and trigger false detection at any attacker-chosen positions, or \textit{appearing attacks}. Such attacks can cause various potential safety hazards such as traffic sign violations, unexpected emergency braking, speeding, etc. Due to such a high potential for practical impacts, these physical-world adversarial attacks on TSR have drawn wide attention across not only the technology community~\cite{whydnn, physicalaednn, aetsr, confusion, streatsignmodifu} but also the general public~\cite{securitynews, bamboozle, hackingstreetsign, fortune, alteringsign}.

Despite such high practical impact potentials, so far existing works generally only considered evaluating the attack effects on academic TSR models, leaving the impacts of these attacks on real-world commercial TSR systems largely unclear. A few recent works tried to understand such commercial TSR system-level impacts, but their evaluation is all limited to one particular vehicle model~\cite{jia2022fooling, sato2023intriguing}, sometimes even an unknown one~\cite{jia2022fooling}, making both the generalizability and representativeness of these evaluation results questionable. This thus raises a critical research question: \textit{Can any of the existing physical-world TSR adversarial attacks achieve a general impact on commercial TSR systems today?}

To answer this critical research question, in this paper, we perform the first large-scale measurement of physical-world adversarial attacks against commercial TSR systems. In this measurement, we focus on hiding attacks as they can most directly impair the function of a commercial TSR product (i.e., by nullifying the TSR function), and test their black-box transfer attack effectiveness against commercial TSR systems, which is the most practical threat model against commercial systems and also the exact setup used by prior works to claim their attack effects on commercial systems~\cite{jia2022fooling, sato2023intriguing}. In total, we were able to include four different commercial vehicle models in this testing, all of which are from the top 15 best-selling vehicle brands in the US \newparts{to ensure high representativeness} (\S\ref{sec:TSR}). \newparts{As estimated later in~\S\ref{sec:commercial_selection}, this setup can be generalizable to at least 33.2\% of the commercial TSR systems sold in the U.S. in 2023, which is significantly improved over prior works (at most 3.8\% or unknown).} We tested all prior works that demonstrated black-box attack transferability in the physical world, which are presumably those having the highest potential to successfully attack commercial systems.

Our testing results reveal that it is actually possible for existing attack works from academia to have a highly-reliable 100\% attack success rate against certain commercial TSR system functionality, which is much higher than expected if compared to the transfer attack success rates reported by the corresponding original papers (e.g., for one such case the reported was $<$20\%). \newparts{Meanwhile,} we do not see generalizability of such attack capabilities over different commercial vehicle models and sign types. Over the entire 30 attack test combinations \newparts{of different sign types, attack methods, surrogate models, and vehicle models}, the vast majority (28/30) do not show any successful attack effects, leading to a 6.67\% overall transfer attack success rate against commercial TSR systems, which is almost a magnitude lower than those reported in the original papers (51.6\% on average).

The much lower-than-expected black-box transfer attack success rate on commercial systems suggests the potential existence of deeper challenges for such attacks to take effect at the TSR system level. Through our investigations, we find that one major factor might be an unexpected \textit{spatial memorization design} that commonly exists in commercial TSR systems today. Specifically, this design exhibits an effect that once a sign is detected, both the detected sign type and the detected location are persistently memorized until the sign's reaction task is finished. Different from methods like object tracking that can only temporally memorize a detection result for a very short time (typically $<$ 3 sec), the spatial memorization we observed will only forget/clear a detection result after the sign's reaction need in the spatial domain is met (e.g., when the vehicle passes the detected sign), regardless of time.

Such a spatial memorization design can significantly impact the success of existing adversarial attacks at the TSR system level. For example, for hiding attacks, to achieve a system-level success in which the TSR system is unable to show the sign display at the sign's reaction task period, the attack has to be continuously successful at \textit{all} possible detection moments that can trigger such memorization before the vehicle passes the sign. For appearing attacks, such an impact on the TSR system-level attack success is the \textit{opposite}: as long as the attack can succeed in \textit{any} of such detection moments, the TSR system-level attack effect can be achieved.

Since such a spatial memorization design commonly existing in commercial TSR systems today may create a significant discrepancy between the TSR model-level attack effect and that at the TSR system level, we further mathematically model its impact on the TSR system-level attack success for both hiding and appearing attacks, which results in new attack success metric designs that can systematically consider the spatial memorization effect. Through both theoretical proof and numerical analysis using these new metric designs, we find that due to spatial memorization, hiding attacks are theoretically harder (if not equally hard) than appearing attacks in achieving TSR system-level attack success. Such an attack hardness gap can be huge ($\geqslant$93.8\% absolute differences in attack success rate values). Meanwhile, due to the lack of consideration of spatial memorization, existing TSR model-level attack success metrics can be highly misleading in judging the TSR system-level attack success, with a potential of having $\sim$50\% absolute attack success rate value differences.

Due to such potential huge differences in judging the TSR system-level attack success, we then use the new attack success metrics to revisit the evaluations, designs, and capabilities of existing attacks in this problem space. These efforts lead to various new findings compared to existing knowledge in this problem space, some of which directly challenge the observations or claims in prior works due to the introduction of the new attack success metrics. For example, we find that while some prior works in this problem space can be claimed as effective using prior success metrics (e.g., with $\sim$50\% to 90\% success rates), when spatial memorization is considered, their success rates can drop significantly to $\leqslant$6.6\% even in white-box attack settings, making it no longer approximate to claim them as effective at the TSR system level. As another example, also due to spatial memorization, we find that the benefits of certain prior attack designs can be seemingly high (e.g., $>$20\% attack success rate increase) using prior metrics, but are actually nearly negligible (e.g., only 1\% increase) at the TSR system level after spatial memorization is considered. The code and data will be made available at our website: \textbf{\url{https://sites.google.com/view/av-ioat-sec/commercial-tsr-test}}.

To sum up, this paper makes the following contributions:
\begin{customitemize}
\item \textit{First large-scale commercial system measurements:} We conduct the first large-scale measurement of physical-world adversarial attacks against commercial TSR systems. Our testing results reveal that although it is possible for existing attack works from academia to have highly reliable (100\%) attack success against certain commercial TSR system functionality, such black-box commercial system attack capabilities are not generalizable, leading to a much lower-than-expected overall black-box transfer attack success rates.

\item \textit{Discovery and analysis of spatial memorization:} We discover a spatial memorization design that commonly exists in today's commercial TSR systems, which can keep memorizing a sign detection result until the sign’s reaction need in the spatial domain is met (e.g., when the vehicle passes the detected sign's position). This discovery is crucial as it is shown to be capable of creating a significant discrepancy between the TSR model-level attack effect and that at the TSR system level.

\item \textit{New attack success metric designs}: We mathematically model the impact of this design on the TSR system-level attack success for both hiding and appearing attacks, resulting in new attack success metric designs that can systematically consider the spatial memorization effect. We then use them to revisit the evaluations, designs, and capabilities of existing attacks in this problem space. 

\item \textit{New observations:} Through the commercial TSR system measurements, new metric designs and analysis, and the revisiting of existing attacks, we uncover a total of 7 novel observations compared to existing knowledge in this problem space, some of which directly challenge the observations or claims in prior works due to the introduction of the new attack success metrics.

\end{customitemize}

\begin{table}
\centering
\small
\tabcolsep 0.045in
\caption{Top 15 leading car brands in the United States based on vehicle sales in 2023~\cite{carsale}. The ones with direct evidence of equipping front windshield cameras for TSR from their official websites or vehicle manuals are marked with check-marks. The vehicle models from 4 out of the 5 highlighted brands below (with bold and underline) are tested in our study (we choose to not directly reveal which four for anonymity purpose).}
\begin{tabular}{clc}
\toprule
Car brand & Sales number & TSR \\
\midrule
Ford & \begin{tikzpicture}[baseline=0.3ex]
    \draw[fill=lightgray] (0,0) rectangle (5.78,0.35); 
    \node[anchor=west, black, font=\scriptsize] at (0.0,0.15) {1,904,038}; 
    \end{tikzpicture} & \cmark\\
    \midrule
 \textbf{\underline{Toyota}} & \begin{tikzpicture}[baseline=0.3ex]
    \draw[fill=lightgray] (0,0) rectangle (5.74,0.35); 
    \node[anchor=west, black, font=\scriptsize] at (0.0,0.15) {1,888,941}; 
    \end{tikzpicture} & \cmark\\
    \midrule
Chevrolet & \begin{tikzpicture}[baseline=0.3ex]
    \draw[fill=lightgray] (0,0) rectangle (5.17,0.35); 
    \node[anchor=west, black, font=\scriptsize] at (0.0,0.15) {1,702,700}; 
    \end{tikzpicture} & \\
    \midrule
Honda & \begin{tikzpicture}[baseline=0.3ex]
    \draw[fill=lightgray] (0,0) rectangle (3.52,0.35); 
    \node[anchor=west, black, font=\scriptsize] at (0.0,0.15) {1,156,591}; 
    \end{tikzpicture} & \cmark\\
    \midrule
\textbf{\underline{Nissan}} & \begin{tikzpicture}[baseline=0.3ex]
    \draw[fill=lightgray] (0,0) rectangle (2.54,0.35); 
    \node[anchor=west, black, font=\scriptsize] at (0.0,0.15) {834,091}; 
    \end{tikzpicture} & \cmark\\
    \midrule
\textbf{\underline{Hyundai}} & \begin{tikzpicture}[baseline=0.3ex]
    \draw[fill=lightgray] (0,0) rectangle (2.42,0.35); 
    \node[anchor=west, black, font=\scriptsize] at (0.0,0.15) {796,506}; 
    \end{tikzpicture} & \cmark\\
    \midrule
Kia & \begin{tikzpicture}[baseline=0.3ex]
    \draw[fill=lightgray] (0,0) rectangle (2.38,0.35); 
    \node[anchor=west, black, font=\scriptsize] at (0.0,0.15) {782,468}; 
    \end{tikzpicture} & \cmark\\
    \midrule
Jeep & \begin{tikzpicture}[baseline=0.3ex]
    \draw[fill=lightgray] (0,0) rectangle (1.95,0.35); 
    \node[anchor=west, black, font=\scriptsize] at (0.0,0.15) {641,166}; 
    \end{tikzpicture} & \cmark\\
    \midrule
Subaru & \begin{tikzpicture}[baseline=0.3ex]
    \draw[fill=lightgray] (0,0) rectangle (1.92,0.35); 
    \node[anchor=west, black, font=\scriptsize] at (0.0,0.15) {632,083}; 
    \end{tikzpicture} & \\
    \midrule
GMC & \begin{tikzpicture}[baseline=0.3ex]
    \draw[fill=lightgray] (0,0) rectangle (1.71,0.35); 
    \node[anchor=west, black, font=\scriptsize] at (0.0,0.15) {563,692}; 
    \end{tikzpicture} & \cmark\\
    \midrule
Ram & \begin{tikzpicture}[baseline=0.3ex]
    \draw[fill=lightgray] (0,0) rectangle (1.64,0.35); 
    \node[anchor=west, black, font=\scriptsize] at (0.0,0.15) {539,477}; 
    \end{tikzpicture} & \cmark\\
    \midrule
\textbf{\underline{Tesla}} & \begin{tikzpicture}[baseline=0.3ex]
    \draw[fill=lightgray] (0,0) rectangle (1.51,0.35); 
    \node[anchor=west, black, font=\scriptsize] at (0.0,0.15) {498,000}; 
    \end{tikzpicture} & \cmark\\
    \midrule
\textbf{\underline{Mazda}} & \begin{tikzpicture}[baseline=0.3ex]
    \draw[fill=lightgray] (0,0) rectangle (1.11,0.35); 
    \node[anchor=west, black, font=\scriptsize] at (0.0,0.15) {365,044}; 
    \end{tikzpicture} & \cmark\\
    \midrule
    BMW & \begin{tikzpicture}[baseline=0.3ex]
    \draw[fill=lightgray] (0,0) rectangle (1.10,0.35); 
    \node[anchor=west, black, font=\scriptsize] at (0.0,0.15) {361,654}; 
    \end{tikzpicture} & \cmark\\
    \midrule
Volkswagen & \begin{tikzpicture}[baseline=0.3ex]
    \draw[fill=lightgray] (0,0) rectangle (1.0,0.35); 
    \node[anchor=west, black, font=\scriptsize] at (0.0,0.15) {329,025}; 
    \end{tikzpicture} & \cmark\\
\bottomrule 
\label{tab:sale}
\end{tabular}
\end{table}

\begin{table}[t]
\tabcolsep 0.06in
\small
    \caption{Existing works with successfully demonstrated traffic sign hiding or appearing attack effects in the physical world using sticker patches or whole-sign posters. HA: Hiding Attack. AA: Appearing Attack. As shown, the commercial system testing aspect is currently under-studied. Highlighted in gray rows are the works that have demonstrated black-box attack transferability in the physical world and thus have the highest potential to successfully attack commercial systems, which are thus the targets of our study \newparts{later in~\S\ref{sec:commercial_testing} and~\S\ref{sec:metric_revisit}}.}
    \centering
    \begin{tabular}{cccccc}

    \toprule
    Existing &  &  &  & Demonstrated & Commercial\\
    works & Year&  HA&AA & transferability? & system testing?\\
    \midrule
    AEFD~\cite{lu2017adversarial} & 2017 & \cmark & & No & None \\
    \rowcolor{gray!30}
    RP$_2$~\cite{eykholt2018physical} & 2018 & \cmark &\cmark & Yes & None \\
    \rowcolor{gray!30}
    SIB~\cite{zhao2019seeing} & 2019 & \cmark &\cmark & Yes & None \\
    \rowcolor{gray!30}
   & &  & & & 1 unknown \\
   \rowcolor{gray!30}
     \multirow{-2}{*}{FTE~\cite{jia2022fooling}}& \multirow{-2}{*}{2022}&  \multirow{-2}{*}{\cmark}& \multirow{-2}{*}{\cmark} & \multirow{-2}{*}{Yes}& vehicle model \\
    SysAdv~\cite{Wang_2023_ICCV} & 2023 & \cmark & & No & None  \\
    \rowcolor{gray!30}
    DM~\cite{sato2023intriguing} & 2024 &   & \cmark& Yes & 1 Tesla model \\    
    \bottomrule
         
    \end{tabular}
    \label{tab:survey}
\end{table}

\section{Background and Related Work}
\label{sec:background}
In the section, we introduce the background for Traffic Sign Recognition (TSR) systems, physical-world adversarial attacks against TSR system, security of autonomous driving (AD) systems, and the threat model for this study.

\subsection{Traffic Sign Recognition (TSR) System}
\label{sec:TSR}

As a key component of Advanced Driver Assistance Systems (ADAS), Traffic Sign Recognition (TSR) system is defined as a system that employs camera sensors to detect road signs, including but not limited to speed limit and STOP signs~\cite{akatsuka1987road, wiki_tsr, kbb_tsr}. Today, this technology is highly prevalent across various vehicle brands to enhance both safety and driving comfort. Table~\ref{tab:sale} shows the top 15 leading car brands in the United States based on vehicle sales in 2023~\cite{carsale}. We surveyed their official websites and vehicle manuals, and were able to find direct evidence of equipping front windshield cameras for TSR for at least 13 out of these 15 brands. 

Recent advancements in Deep Neural Networks (DNNs) have propelled significant progress in various domains, including TSR systems, which now increasingly rely on DNN-based methodologies for real-time object detection~\cite{sato2024invisible, jia2022fooling, zhao2019seeing, almutairy2019arts}. These systems process camera sensor data through DNN-based object detectors to identify road signs efficiently. Current state-of-the-art object detection models are categorized into two primary types: one-stage and two-stage detectors~\cite{zou2023object}. One-stage detectors, such as YOLO~\cite{redmon2018yolov3}, are celebrated for their rapid detection capabilities. Conversely, two-stage detectors, exemplified by Faster R-CNN~\cite{ren2015faster}, are noted for their exceptional accuracy. Prior research~\cite{zhao2019seeing, Wang_2023_ICCV, sato2024invisible} focusing on the security aspects of TSR systems has examined models from both categories for evaluation comprehensiveness. In line with these research, our analysis also encompasses object detectors from both categories, aiming to provide a comprehensive assessment of TSR systems security.

\subsection{Physical-World Adversarial Attacks against TSR}
\label{sec:TSR_attack}

DNN models today are shown to be generally vulnerable to adversarial examples (or \textit{adversarial attacks})~\cite{szegedy:iclr:2014, goodfellow:fsgm,carlini:cw,papernot:jsma,madry:towards,pei2017deepxplore, zhang2020interpretable, cao2021invisible}. Such vulnerabilities are especially extensively studied and demonstrated in the vision domain~\cite{szegedy:iclr:2014, goodfellow:fsgm,carlini:cw,papernot:jsma,madry:towards,pei2017deepxplore, zhang2020interpretable, jia2020fooling, ma2023slowtrack, sato2021wip, jiang2019black, Wang_2023_ICCV, sato2021dirty}. Due to the increasing real-world penetration of TSR systems and their fundamental reliance on the camera inputs, TSR models soon became a natural target of adversarial attack research, including many that were able to achieve particularly high realism with successfully demonstrated attack effect in the physical world~\cite{eykholt2018physical, lu2017adversarial, zhao2019seeing, chen2018shapeshifter, nassi2020phantom, zhu2023tpatch, Wang_2023_ICCV, sato2023intriguing, sato2024invisible, lovisotto2021slap, duan2021adversarial, zhong2022shadows, ma2024controlloc}.

Among them, the most representative and also the most widely-exploited attack vectors are \textit{physical patches/posters}, e.g., by physically printing attack patterns on \textit{sticker patches} and attaching them to the legitimate traffic sign surface~\cite{eykholt2018physical, Wang_2023_ICCV, zhao2019seeing, hedorpatch}, or on \textit{whole-sign posters} that replace or spoof the entire traffic sign surface~\cite{lu2017adversarial,  jia2022fooling, zhao2019seeing, eykholt2018physical, sato2023intriguing}. These attacks are low-cost, highly deployable, and demonstrated capable of causing various severe attack effects, most notably (1) making critical legitimate traffic signs undetectable, most representatively \textit{hiding attack}, or HA~\cite{eykholt2018physical, zhao2019seeing, Wang_2023_ICCV, jia2022fooling}; and (2) triggering false detection at any attacker-chosen positions, most representatively \textit{appearing attack}, or AA~\cite{eykholt2018physical, zhao2019seeing, jia2022fooling, sato2023intriguing}. For drivers who are relying on such a driver assistance function, or higher-autonomy AD systems that try to automatically react to real-time TSR results (e.g., in Tesla~\cite{tesla-traffic-sign-control}), such attacks, especially the hiding ones, can directly impair the TSR function and cause various potential safety hazards such as traffic sign violations, unexpected emergency braking, speeding, etc. Due to such a high potential for practical impacts, these physical-world adversarial attacks on TSR have drawn wide attention in not only the technology community~\cite{whydnn, physicalaednn, aetsr, confusion, streatsignmodifu} but also the general public~\cite{securitynews, bamboozle, hackingstreetsign, fortune, alteringsign}.

Despite such a high practical impact potential, so far these works generally only considered evaluating the attack effects on academic TSR models, leaving the impacts of these attacks on real-world commercial TSR systems largely unclear.
Table~\ref{tab:survey} summarized all the prior works so far that were able to successfully demonstrate using physical patches/posters to trigger HA or AA attack effects in the physical world. As shown, although many of them have demonstrated the attack transferability across academic models, which could be viewed as an indicator of high potential black-box attack capability against commercial systems, very few have actually evaluated the attacks against real commercial vehicle systems. For the only two works that have done so, the evaluation is limited to one particular vehicle model (for one of them it is even an unknown vehicle model)~\cite{jia2022fooling, sato2023intriguing}, making both the generalizability and representativeness of these evaluation results questionable. 

The observations above thus raise a critical research question: \textit{Can any of these existing physical-world TSR adversarial attacks achieve a general impact on commercial TSR systems today?} In this work, we thus aim to systematically answer this critical research question by performing the first large-scale testing of representative existing attacks against commercial TSR systems of top popularity on the consumer market today. Leveraging the observations and insights from the testing, we also further systematically revisit existing evaluation metrics and attack designs in this problem space.

\subsection{Security of Autonomous Driving (AD) systems}
\label{sec:ad_security}
Due to the fundamental reliance of AD systems on environmental sensing, prior works have extensively investigated sensor attacks within the AD context. These include spoofing or jamming attacks targeting cameras~\cite{yan2016can, nassi2020phantom, sayles2021invisible, kohler2021they}, LiDAR~\cite{cao2019adversarial, shin2017illusion, cao2023you, jin2023pla}, RADAR~\cite{yan2016can}, etc., highlighting vulnerabilities at the sensor level. In contrast, our study focuses on the vulnerabilities at the autonomous AI algorithm level, specifically targeting the TSR functions in AD systems, which is highly crucial for safe and correct driving automation. While the existing body of literature covers security aspects of various components such as camera object detection~\cite{eykholt2018physical, chen2018shapeshifter, zhao2019seeing, jia2022fooling, cao2021invisible, Wang_2023_ICCV}, object tracking~\cite{jia2020fooling, ma2023slowtrack}, lane detection~\cite{sato2021dirty}, and end-to-end AD systems~\cite{pei2017deepxplore, tian2018deeptest}, there is a notable gap in large-scale studies on their effectiveness in real-world commercial AD systems. In this work, we thus aim to bridge this critical research gap by conducting the first large-scale measurements of physical-world adversarial attacks on commercial TSR systems, which not only expands the understanding of such security vulnerabilities in real-world AD systems but also provides various new insights into the design and evaluation of existing works in this problem space.

\subsection{Threat Model}
\label{sec:threatmodel}

To understand the impacts of these existing physical-world TSR attack works from the commercial systems perspective, we consider the most realistic \textit{transferability-based black-box} threat model, i.e., generating the adversarial attack pattern using publicly-accessible surrogate TSR models and then applying it to the attack-targeted TSR systems with unknown model parameters and architectures. We choose this because (1) as with most commercial products, commercial TSR systems, especially the most popular ones today in Table~\ref{tab:sale}, are by default closed-source to the public and so far there is no effective approach to generally reverse-engineer them; and (2) this is also the setup used by existing physical-world TSR attack works to claim their attack effects on commercial systems~\cite{jia2022fooling, sato2023intriguing}. For attack goals, we consider both traffic sign hiding and appearing attacks as highlighted in \S\ref{sec:TSR_attack}, with a more specific focus on the hiding attack side as it can most severely impair a commercial TSR product by completely nullifying a TSR system's functionality. For attack vectors, we follow the most representative and practical physical patch/poster attack vectors (\S\ref{sec:TSR_attack}). Specifically, we use sticker patches for hiding attacks (HA) and whole-sign posters for appearing attacks (AA), which are the most practically-deployable attack vectors on both sides~\cite{eykholt2018physical, zhao2019seeing, jia2022fooling, Wang_2023_ICCV, sato2023intriguing}.



\section{Large-Scale Commercial TSR Systems Testing and Observations}
\label{sec:commercial_testing}

In this section, we report our efforts on the first large-scale testing of existing physical-world TSR model attacks on commercial TSR systems. In this testing, we specifically focus on the hiding attacks (HA) as they can make critical traffic signs undetectable and thus most directly impair the function of a commercial TSR product (\S\ref{sec:threatmodel}). We first detail the experimental setup and then report our key observations.

\subsection{Attack Setup}
\label{sec:setup}

{\bf Traffic Sign Selection.} Our study considers two types of traffic signs: the STOP sign and the 25 mph speed limit sign. We choose these two types of signs since (1) STOP and speed limit signs are the most popular targets in prior works for demonstrating the adversarial attack effects in the physical world~\cite{lu2017no, jia2022fooling, eykholt2018physical, zhao2019seeing, xue2021naturalae, lu2017adversarial, chen2018shapeshifter, sato2023intriguing, Wang_2023_ICCV, sato2024invisible}; and (2) both are highly safety-critical as missing STOP signs can lead to intersection collisions and 25 mph speed limit sign is usually for residential or school districts where children can be outside or crossing the street~\cite{speedbasics}. \newparts{Meanwhile, these two sign types are also the only two types of signs with demonstrated physical-world black-box attack transferability in prior works~\cite{eykholt2018physical, zhao2019seeing, jia2022fooling}.}

\begin{table}[t]
\small
    \caption{Attack success rates for our reproduced RP$_2$, SIB, and FTE attacks compared to those in the original papers following the same surrogate model and attack distance setups.}
    \centering
    \begin{tabular}{cccc}

    \toprule
    Attack success rate & RP$_2$~\cite{eykholt2018physical} & SIB~\cite{zhao2019seeing} & FTE~\cite{jia2022fooling} \\
    \midrule
    Reported by the papers & 63.5\% & 60.5\% & 98.8\%\\
    Our reproduced attacks & 60.5\% & 84.5\% & 98.7\%\\

    \bottomrule
         
    \end{tabular}
    \label{tab:reproduction}
\end{table}

{\bf Selected Attacks.} As shown in Table~\ref{tab:survey}, there are three prior works so far that were able to demonstrate black-box attack transferability for the hiding attack effect in the physical world: RP$_2$\cite{eykholt2018physical}, SIB~\cite{zhao2019seeing}, and FTE~\cite{jia2022fooling}. Specifically, FTE was able to demonstrate this against a commercial vehicle model, despite an unknown one~\cite{jia2022fooling}. These three works have the highest potential to successfully attack commercial systems; thus, we aim to include all of them in our testing. Unfortunately, at this point, none of these three works have open-sourced their methods. Thus, we have to reproduce them. We tried our best to reproduce them, which included both following their papers closely and consulting with the original paper authors for all three, and the reproduced attack success rates are shown in Table~\ref{tab:reproduction}. Note that there are some performance discrepancies between our reproduced version and the original one. These discrepancies can be due to various factors, most likely the differences in the experimental setups such as the hyperparameters used in attack generation, the real-world data collected for physical-world robustness, and realizability optimization, etc. Despite our best efforts to replicate their original experimental setup, including consulting the original paper authors, it is fundamentally impossible for us to exactly replicate their setup due to the lack of open-sourcing in these prior works.

{\bf Surrogate Model and Dataset Selection.} 
As detailed in~\S\ref{sec:threatmodel}, in this testing, we adopt the most realistic transferability-based black-box threat model. To achieve this, we carefully select surrogate models and datasets, based on the selection by previous research in this field~\cite{sato2024invisible, Wang_2023_ICCV, jia2022fooling}. Specifically, we utilize the Microsoft COCO dataset~\cite{lin2014microsoft} to study adversarial attacks on STOP signs and the ARTS dataset~\cite{almutairy2019arts} for speed limit signs. Our surrogate model selection covers both one-stage and two-stage TSR model designs (S\ref{sec:TSR}) to increase the chance of successful transfer attacks. Specifically, we choose YOLO v5 (denoted as \textit{Y5}) for the one-stage model and Faster RCNN (denoted as \textit{FR}) as the two-stage one, both of which are from the most widely-used model families in prior works~\cite{jia2022fooling, zhao2019seeing, eykholt2018physical, chen2018shapeshifter, zhu2023tpatch}. In particular, Y5 is also the one that has been used to demonstrate a successful transfer attack to a commercial vehicle model~\cite{jia2022fooling}. For Faster RCNN, we adopt the latest official PyTorch implementation that uses the ResNet-50-FPN-V2 backbone~\cite{li2021benchmarking}. For the Microsoft COCO dataset, the models are obtained directly from the Y5 official website~\cite{glenn_jocher_2020_4154370} and the PyTorch models~\cite{torchmodel}. For the ARTS dataset, we conduct our own model training. The benign performances of these models have an mAP (mean Average Precision) of 0.831 for Y5 and an mAP of 0.871 for FR, which are consistent with those reported in prior research~\cite{sato2024invisible}.

\begin{figure}[t]
    \footnotesize
      \centering
        \includegraphics[width=0.9\linewidth]{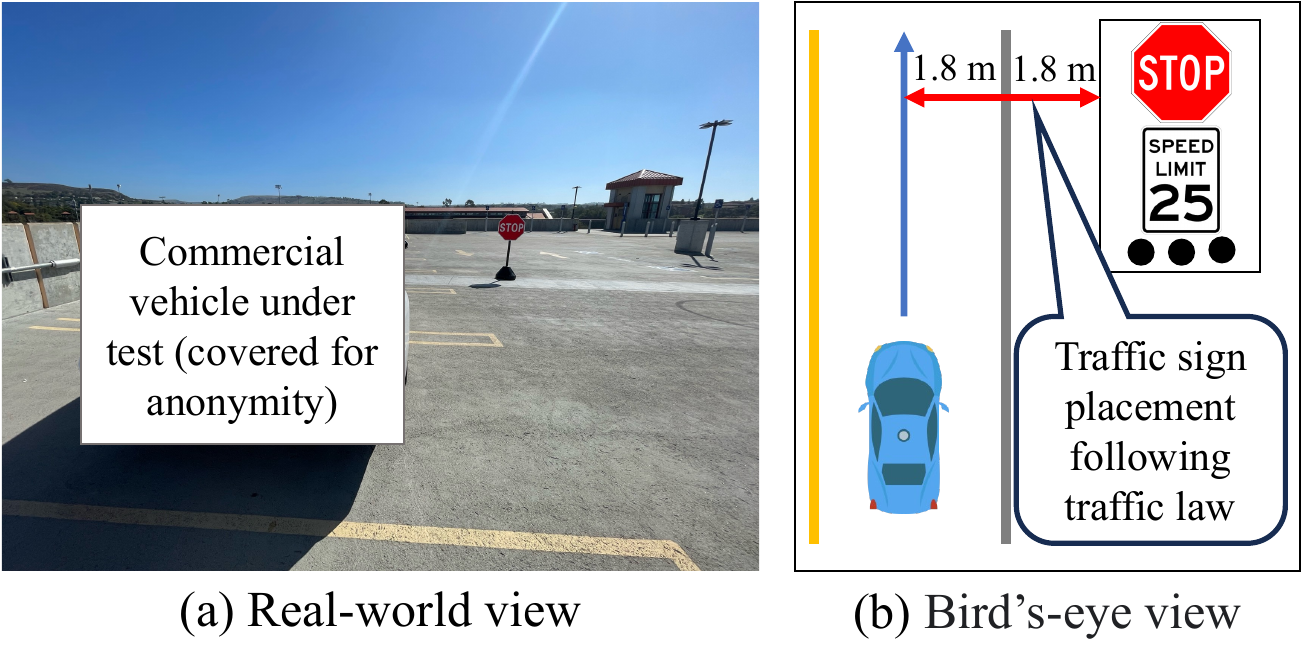}  
    \caption{\newparts{Experiment setup for commercial TSR system testing. We cover the vehicle in the photo for anonymity purpose.}}
        \label{fig:test_env_visual}
\end{figure}

{\bf \newparts{Test Environment Setups}.} 
Our experiments are performed outdoors during sunny afternoons between 1 pm and 4 pm, to simulate the most common real-world attack scenarios. To maintain consistent testing conditions, we measure the ambient light levels using a light meter, ensuring that all tests are conducted within a light range of 25,000 to 30,000 lux. 
Visual representations of the real-world environment and its bird's-eye view illustration are provided in Fig.~\ref{fig:test_env_visual}. 
We select a spacious rooftop parking structure as the location for these experiments ensuring no presence of other vehicles or humans to maintain safety. The placement of traffic signs is carefully designed to follow traffic laws in the U.S. as outlined in previous studies~\cite{Wang_2023_ICCV}. \newparts{For the testing distance, we carefully set the start point to be farther than the detection distance in the benign case for each vehicle model ($\sim$50 meters). For the testing speed, we test at the maximum-allowed speed limit of our rooftop parking structure (5 mph) to ensure safety.}

\begin{table*}[t]
\small
    \caption{Four of the five commercial vehicle models below are tested in our study (denoted as C1 to C4 in Table~\ref{tab:vehicle_info}). Each model has functions to detect a STOP sign, speed limit signs, or both. \newparts{We choose not to directly reveal the exact models for C1 to C4 for anonymity purpose. Note that this is the least anonymization to protect the affected companies (i.e., only 1 confusing vehicle model), and this is also part of the agreement with the companies during our responsible vulnerability disclosure.}}
    \centering
    \begin{tabular}{ccccc}

    \toprule
    Tesla Model 3 2023& Toyota Camry 2023& Nissan Sentra 2023 & Mazda CX-30 2023& Hyundai Tucson 2024\\

    \midrule
    \\
    \\
    \multirow{-3}{*}{\includegraphics[width=0.12\linewidth]{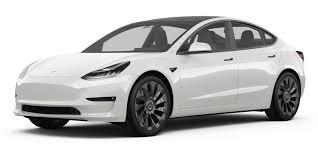}}   & \multirow{-3.1}{*}{\includegraphics[width=0.12\linewidth]{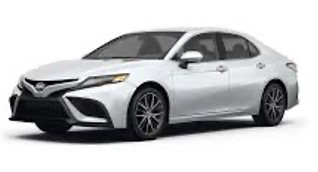}}  & \multirow{-3}{*}{\includegraphics[width=0.115\linewidth]{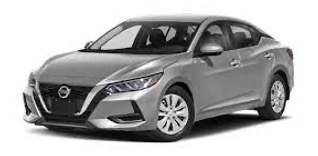}} & \multirow{-2.9}{*}{\includegraphics[width=0.095\linewidth]{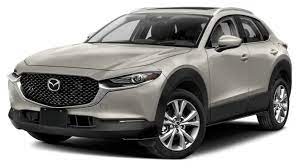}} & \multirow{-3.2}{*}{\includegraphics[width=0.11\linewidth]{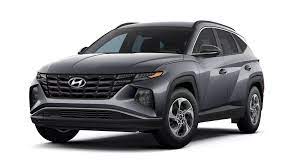}}\\

    \bottomrule
         
    \end{tabular}
    \label{tab:car_company}
\end{table*}

\begin{table}[t]
\small
    \caption{TSR functions of the four vehicle models tested in our measurement study. These four models are among the five in Table~\ref{tab:car_company}. We choose to not directly reveal the exact vehicle brands and models for anonymity purpose.}
    \centering
    \begin{tabular}{ccc}

    \toprule
     &  \multicolumn{2}{c}{TSR functionality} \\
    \cmidrule(lr){2-3}
    {Vehicle model}  & STOP sign & Speed limit sign \\
    \midrule
    Car 1  (denote as C1)  &  {\large \cmark} &   {\large \xmark}\\
    Car 2  (denote as C2)  &   {\large \cmark} &   {\large \cmark}\\
    Car 3  (denote as C3)   &  {\large \xmark} &   {\large \cmark}\\
 
    Car 4 (denote as C4) &  {\large \xmark} &   {\large \cmark}\\

    \bottomrule
         
    \end{tabular}
    \label{tab:vehicle_info}
\end{table}

\subsection{Commercial Systems Under Test and Metric}
\label{sec:commercial_selection}

{\bf Commercial Systems Under Test.} We were able to include 4 different vehicle models in this testing through borrowing or renting. These four vehicle models are among the five in Table~\ref{tab:car_company}. As shown, all of them are among the top 15 popular vehicle brands in the US based on vehicle sales (Table~\ref{tab:sale}) and all of them are from the most recent model years, either 2023 or 2024. Note that we choose to not directly reveal the exact vehicle brands and models for anonymity purpose; \newparts{this is already the least anonymization to protect the affected companies (i.e., only 1 confusing vehicle model), and this is also part of the agreement with the companies during our responsible vulnerability disclosure.} In the paper, we denote the four vehicle models we tested as C1 to C4. Table~\ref{tab:vehicle_info} shows the TSR functionality support we found for C1 to C4 using benign STOP and speed limit signs. As shown, two of them can support STOP sign detection, while three of them can support speed limit sign detection. In particular, C2 can support both.

\newparts{\textbf{Generalizability of the systems under test.} While the rankings of the vehicle sales in Table~\ref{tab:sale} have quantifiably shown the representativeness of the vehicle models under test, it is better if we can further quantify the generalizability of this tested system setup. 
To achieve this, we use the market share of the tested vehicle models as an estimate. Specifically, in 2023 the U.S. automotive industry sold around 15.6 million vehicles~\cite{ussold}. Based on the data in Table~\ref{tab:sale}, at most 13.3 million of these 15.6 million vehicles are TSR-equipped. The total sales of the 5 possible vehicle brands in our test account for 33.2\% of such upper-bound number of TSR-equipped vehicles sold in 2023. Thus, our current testing results can be estimated as generalizable to at least 33.2\% of the commercial TSR systems sold in the U.S. in 2023. Using this estimation, our commercial TSR systems testing results can also be shown to be much more generalizable than prior works in Table~\ref{tab:survey} (i.e., at most 3.8\% for~\cite{sato2023intriguing} and unknown for~\cite{jia2022fooling}).}

\textbf{TSR System-Level Attack Success Metric.}
\label{sec:metric_naive}
In prior works, the TSR adversarial attack success rates are generally calculated by TSR model-level metrics, i.e., first determining the attack success at TSR model output level per frame and then aggregating the per-frame results over one or multiple distance ranges~\cite{zhao2019seeing, eykholt2018physical, shen2022sok, lovisotto2021slap, zhu2023tpatch, Wang_2023_ICCV, sato2024invisible} or a certain number of consecutive frames~\cite{zhao2019seeing, shen2022sok}. However, we find that the TSR systems in commercial vehicle models today do not generally show real-time traffic sign detection results to end users; instead, the duration and timing of the detection result display are more generally based on the \textit{system-level needs} for different sign types. For example, for speed limit signs, we find that all the vehicle models supporting them (C2 to C4) do not immediately show the speed limit sign detection results when the sign is actually detected; instead, the detection results will only be on display \textit{after} the vehicle passes the sign \newparts{(more precisely, always when the vehicle body is halfway past the sign in our experiments)}. This TSR system-level design aligns with system-level needs for such traffic sign detection functionality, as a newly-detected speed limit should be applied \textit{after} the vehicle drives past the physical location of the sign~\cite{trafficlaw1}. \newparts{This design can also be viewed as reasonable if we consider the design of the speed limit sign display is to indicate the speed limit of the current road segment, so it is indeed correct to only show the new speed limit after it enters the corresponding road segment. If it shows the new speed before that, the car can be speeding before it enters a road segment with a higher speed limit.}

For the STOP sign, we have a similar observation: although it is different from speed limit signs in that it should be displayed before the vehicle passes the sign, we find that in C2 once the sign is detected, the TSR system will keep having the sign display instead of showing the real-time detection results until it passes the sign (this is also the spatial memorization effect that we will investigate more later in~\S\ref{sec:spatial_memo}). This again aligns well with the system-level needs, as a detected STOP sign should take effect until it is passed.

Due to these observations, we need to use an attack success metric defined at the TSR system level to most generally and practically-meaningfully capture the impacts of adversarial attacks on commercial TSR systems. To this end, we thus define the TSR system-level attack success per each \textit{traffic sign reaction task} on the TSR system user side, i.e., when the TSR system user needs to react to a sign, if the TSR system is able to correctly display the sign, the attack fails; otherwise, the attack succeeds. For example, for speed limit sign, the attack success is judged by whether the system can have the sign displayed at the time when the vehicle passes the sign, while for STOP sign it is judged by whether the system can have the sign displayed before the vehicle passes the sign.

\begin{table*}[t]
\small
    \caption{Commercial TSR systems testing results against vehicle model C1 to C4 with comparisons to the black-box transfer attack success rates reported by the original papers. The testings for each benign or attack setup are repeated 3 times.
    }
    \centering
    \begin{tabular}{ccccccccc}

    \toprule
    & & & C1 & \multicolumn{2}{c}{C2} & C3& C4&\\
    \cmidrule(lr){4-4}
    \cmidrule(lr){5-6}
    \cmidrule(lr){7-7}
    \cmidrule(lr){8-8}
    & \multirow{-2}{*}{\shortstack{Original paper \\ transferability}}& \multirow{-2}{*}{\shortstack{Surrogate \\ model}}& STOP & STOP & Speed limit& Speed limit& Speed limit& \multirow{-2}{*}{Ave.}\\
    \midrule
    \midrule
    \multicolumn{3}{c}{Benign traffic sign} & \textbf{100\% (3/3)} & \textbf{100\% (3/3)} & \textbf{100\% (3/3)} & \textbf{100\% (3/3)} & \textbf{100\% (3/3)} & \textbf{100\%} \\
    \midrule
    \midrule
    & & Y5 & 0\% (0/3) & 0\% (0/3) & 0\% (0/3) & 0\% (0/3) & 0\% (0/3) & 0\%\\
    \multirow{-2}{*}{RP$_2$} & \multirow{-2}{*}{18.9\%} & FR & 0\% (0/3) & \textbf{100\% (3/3)}& 0\% (0/3) & 0\% (0/3) & 0\% (0/3) & \textbf{20\%} \\
    \midrule
    & & Y5 & 0\% (0/3) & \textbf{100\% (3/3)} & 0\% (0/3) & 0\% (0/3) & 0\% (0/3) & \textbf{20\%} \\
    \multirow{-2}{*}{SIB} & \multirow{-2}{*}{46.1\%} & FR & 0\% (0/3) & 0\% (0/3) & 0\% (0/3) & 0\% (0/3) & 0\% (0/3) & 0\% \\
    \midrule
    & & Y5 & 0\% (0/3) & 0\% (0/3) & 0\% (0/3) & 0\% (0/3) & 0\% (0/3) & 0\% \\
    \multirow{-2}{*}{FTE} & \multirow{-2}{*}{89.8\%} & FR & 0\% (0/3) & 0\% (0/3) & 0\% (0/3) & 0\% (0/3) & 0\% (0/3) & 0\% \\
    \midrule
    Ave. over all attacks & 51.6\% & & 0\% & \textbf{33.3\%} & 0\% & 0\% & 0\%& \textbf{6.67\%} \\
    
    \bottomrule
         
    \end{tabular}
\vspace{0.4cm}
    \label{tab:results_ss}
\end{table*}

\subsection{Testing Results and Observations}
\label{sec:commercial_res}

\begin{figure}[t]
    \footnotesize
      \centering
        \includegraphics[width=0.8\linewidth]{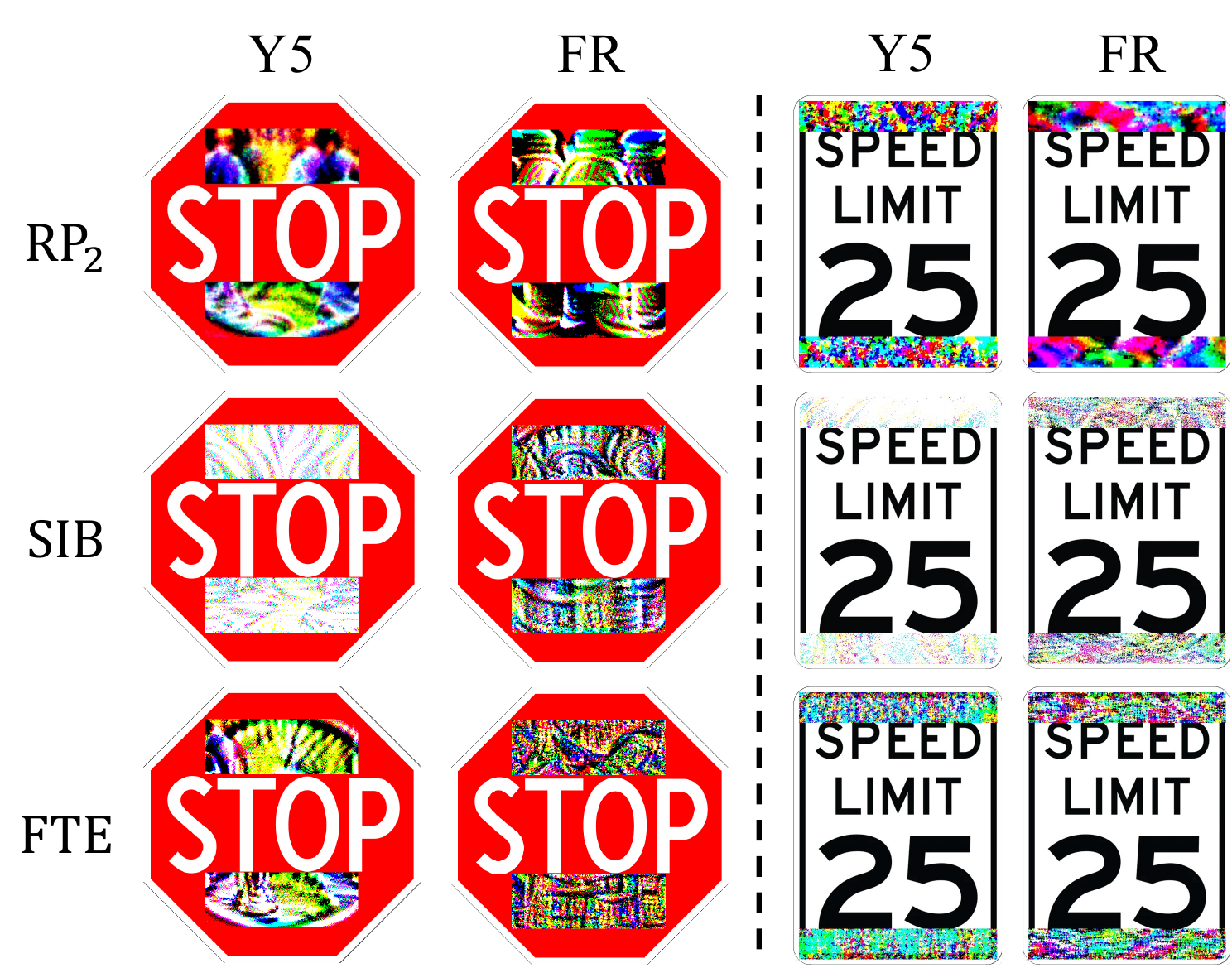} 
    \vspace{0.2cm}
    \caption{Visualisation of the hiding attacks (HA) generated for STOP and speed limit signs, which are used in our commercial TSR systems testing. They are generated by the three most promising prior works (RP$_2$~\cite{eykholt2018physical}, SIB~\cite{zhao2019seeing}, FTE~\cite{jia2022fooling}) using surrogate models of both representative one-stage and two-stage TSR model designs.}
        \label{fig:sign_visual}
\end{figure}

{\bf Overall Testing Results.} Table~\ref{tab:results_ss} summarizes the overall testing results and the reproduced attack visualization is in Fig.~\ref{fig:sign_visual}. As shown, there are 30 attack test combinations in total, each for one combination of the 2 sign types, 3 attack methods, 2 surrogate models, and 4 vehicle models. For each benign and attack setup, we repeat the testing three times. As shown, we are indeed able to find attack setups in which existing physical-world adversarial attack works from academia can reliably work on a certain commercial TSR system, more specifically the RP$_2$ attack using FR as the surrogate model, and the SIB attack using Y5 as the surrogate model. In these setups, the attack can \textit{always} succeed over the three runs, leading to a 100\% success rate. Such a high black-box transfer attack effectiveness is even higher than expected as the transfer attack success rates are actually less than 50\% for both and even less than 20\% for RP$_2$. This suggests that for certain commercial TSR systems, although from top brands in the US (Table~\ref{tab:sale}), their TSR functionality can actually be much more vulnerable than academic TSR models under black-box transfer attacks. Interestingly, both successful attack setups are against C2, the only vehicle model among the four that can support both STOP and speed limit signs. We have already performed responsible vulnerability disclosure to the C2 manufacturer to report these.

While there \newparts{do} exist successful attack cases, we do not see generalizability of such attack effects over the entire testing results. Over the entire 30 attack test combinations, the vast majority (28/30) do not show any successful attack effects, leading to a 6.67\% overall transfer attack success rate against commercial systems. As shown in the table, this is almost \textit{a magnitude lower} than those reported in the original papers~\cite{eykholt2018physical, zhao2019seeing, jia2022fooling} (51.6\% on average). Even for these two successful attack setups, the attack effect is limited to the STOP sign detection and cannot even generalize to the speed limit sign detection for the same vehicle model (C2). \newparts{This could be due to the need to customize the detection accuracy for certain sign types because of real-world deployment or customer needs.} Interestingly, although FTE was able to demonstrate a successful transfer attack against a commercial vehicle model in its paper, we were not able to find any successful attack results against any of the four commercial vehicle models in our tests. This further reveals the lack of generalizability of the reported commercial TSR system attack success in the original FTE paper, which cannot be revealed without the large-scale commercial system testing efforts in this paper.

\newparts{Note that we have performed statistical testing for the results above to understand their statistical significance. While the overall transfer attack success rate against commercial systems over all 30 attack test combinations (6.67\%) is statistically significant, the statistical significance for the result of each test combination cannot be calculated since the variance for each is 0 (all failure or all success as shown in Table~\ref{tab:results_ss}). While it may be possible that some variance can appear if we significantly increase the number of attempts per test combination (e.g., to over 30) to make statistical significance calculable, we cannot afford this due to inherent limitation for any outdoor vehicle testing setups. Note that our current setup is already scientifically more rigorous than all prior works in this as they only tried once for each attack test. More detailed discussions are in~\S\ref{sec:limitation}.}

\begin{mtbox}{Observation\,\stepcounter{countobs}\arabic{countobs}: }
\refstepcounter{refcounter}\label{obs1} It is in fact possible for existing physical-world adversarial attack works from academia to have highly reliable (100\%) attack success against certain commercial TSR system function in practice. However, such black-box commercial system attack capability is currently not generalizable over different representative commercial system models and sign types. Overall, the black-box transfer attack success rate on commercial systems \newparts{(at least on our setup that can account for at least 33.2\% of commercial TSR systems sold in the U.S. in 2023, as estimated in~\S\ref{sec:commercial_selection})} is much lower than that on academic models in prior works.
\end{mtbox}

\begin{table*}[t]
\tabcolsep 0.04in
\small
    \caption{TSR detection result memorization rates when we hide the sign for a long time (20-60 seconds) after a short sign display time (1-3 seconds). The experiment setup is described in Fig.~\ref{fig:spation_visual}. As shown, three out of the four vehicle models exhibit a spatial memorization design, i.e., keeping memorizing a sign detection result until the sign’s reaction need in the spatial domain is met (e.g., when the vehicle passes the detected sign), regardless of time. Notice the observed much longer memorization time (60 sec) than that from typical temporal memorization designs such as object tracking (typically $<$3 sec~\cite{zhu2018online, shen2022sok, Wang_2023_ICCV, jia2020fooling}).
    }
    \centering
    \begin{tabular}{cccccccc}

    \toprule
     &  &\multicolumn{6}{c}{Sign disappearing time after the short sign display}\\
     \cmidrule(lr){3-8}
     &  &\multicolumn{3}{c}{STOP sign} &\multicolumn{3}{c}{Speed limit sign}\\
     \cmidrule(lr){3-5}
     \cmidrule(lr){6-8}
    \multirow{-4}{*}{Vehicle model} & \multirow{-4}{*}{\shortstack{Sign\\display\\time}}
    
    & 20 sec & 40 sec & 60 sec & 20 sec & 40 sec & 60 sec\\
    \midrule
    & 1 sec & 0\% (0/3) & 0\% (0/3) & 0\% (0/3) & - & - & -\\
   \multirow{-2}{*}{C1} & 3 sec & 0\% (0/3) & 0\% (0/3) & 0\% (0/3) & - & - & -\\
   \midrule
   & 1 sec & 100\% (3/3) & 100\% (3/3) & 100\% (3/3) & 100\% (3/3) & 100\% (3/3) & 100\% (3/3) \\
   \multirow{-2}{*}{C2} & 3 sec & 100\% (3/3) & 100\% (3/3) & 100\% (3/3) & 100\% (3/3) & 100\% (3/3) & 100\% (3/3)\\
   \midrule
   & 1 sec & - & - & - & 100\% (3/3) & 100\% (3/3) & 100\% (3/3)\\
   \multirow{-2}{*}{C3, C4} & 3 sec & - & - & - & 100\% (3/3) & 100\% (3/3) & 100\% (3/3)\\
    \bottomrule
         
    \end{tabular}
    \vspace{0.4cm}

    \label{tab:spatial_tracking}
\end{table*}

\textbf{Spatial Memorization of TSR Results.}
\label{sec:spatial_memo}
The much lower-than-expected black-box transfer attack success rate on representative commercial systems suggests the potential existence of deeper challenges for such attacks to take effect at the TSR system level. Through our investigations, one major factor might be an unexpected \textit{spatial memorization design} that commonly exists in commercial TSR systems today. Specifically, this design exhibits an effect that once a traffic sign is detected, both the detected sign type and the detected location are persistently memorized until the sign's reaction task is finished. Different from simple object tracking that can only temporarily memorize a detection result for a very short time (typically at most 3 seconds~\cite{zhu2018online, shen2022sok, Wang_2023_ICCV, jia2020fooling}), the spatial memorization we observed will only forget/clear a detection result after the sign's reaction need in the spatial domain is met (e.g., when the vehicle spatially passes the position of a detection STOP sign or speed limit sign), regardless of time.

\begin{figure}[t]
    \footnotesize
      \centering
        \includegraphics[width=\linewidth]{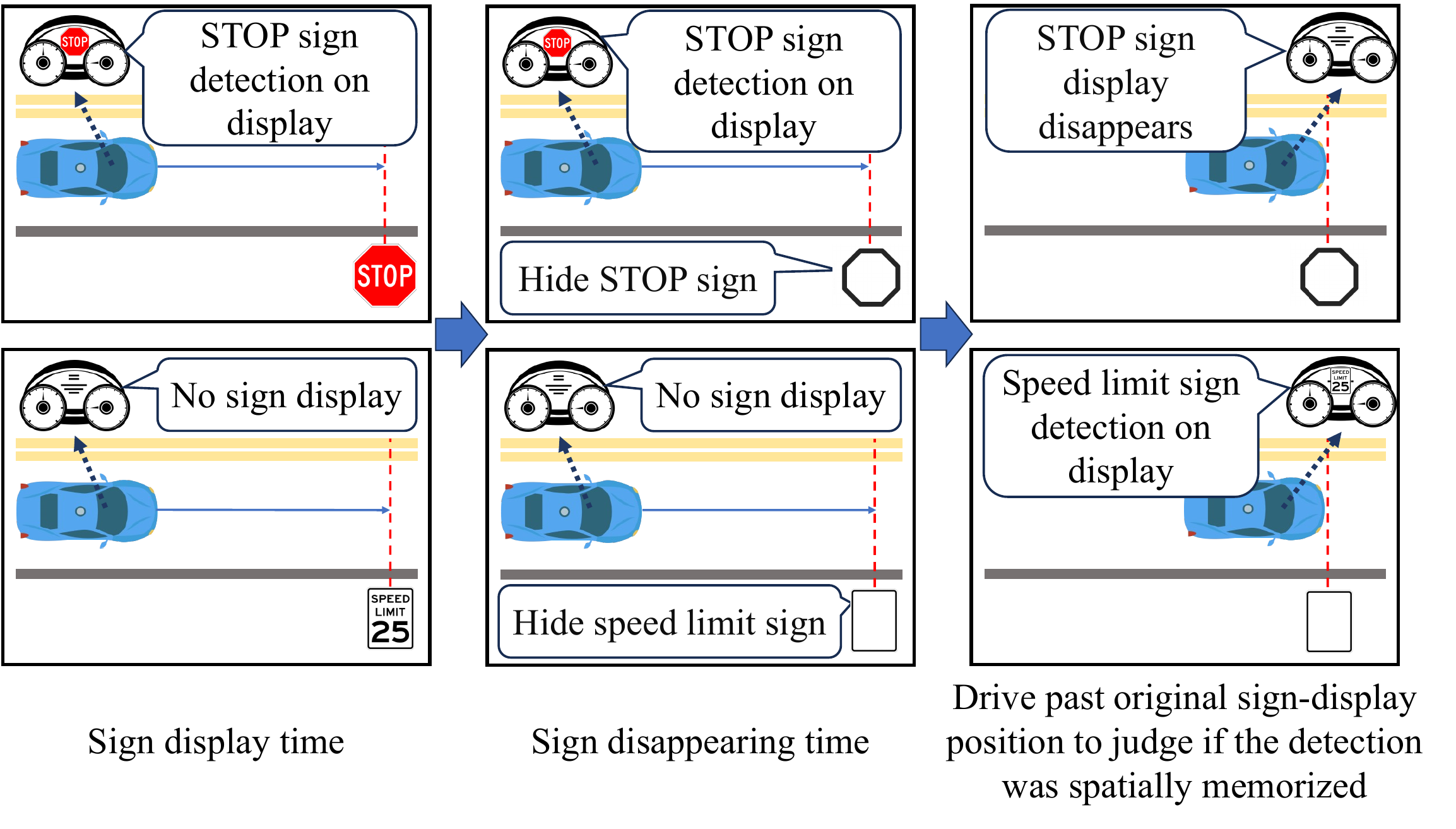} 
    \caption{Experimental setup for our investigation into the spatial memorization design in commercial TSR systems. As shown, we first show the sign to the vehicle for a short time (\textit{sign display time}), and hide the sign and wait for a certain time duration (\textit{sign disappearing time}). After that, we drive the vehicle past the original sign-display position to measure whether the sign detection result is spatially memorized.}
        \label{fig:spation_visual}
\end{figure}

Table~\ref{tab:spatial_tracking} shows our experimental investigation of this design in the four commercial vehicle models. Fig.~\ref{fig:spation_visual} illustrates the experimental setup. As shown, in the experiments we first keep the tested vehicle stationary and show the traffic sign on the roadside in front of the vehicle for 1 to 3 seconds (\textit{sign display time}. Then, we hide the traffic sign and wait for 20 to 60 seconds (\textit{sign disappearing time}). Then, we test whether the sign detection result triggered at the sign display time is still memorized by the TSR system after the sign has disappeared for a certain time by driving the vehicle past the original sign-display position. For the STOP sign, the memorization is judged by whether the sign display disappears after driving past the original sign-display position, and for the speed limit sign, this is judged by whether the sign display appears after driving past the original sign-display position. As shown, for three out of the four vehicle models (C2 to C4), the sign detection result can retain even after the sign has already disappeared for 60 seconds, which is way longer than the typical temporal memorization time from object tracking (3 seconds~\cite{zhu2018online, jia2020fooling}), and will only be cleared/forgotten when the vehicle passes the position of the detected sign. 

Such a spatial memorization design can significantly impact how we judge the adversarial attack effect at the TSR system level. For example, for hiding attacks, to achieve a system-level success in which the TSR system is unable to show the sign display at the sign's reaction task period, the attack has to be continuously successful at all possible detection moments that can trigger such memorization before the vehicle passes the sign. As shown in our experiments in Table~\ref{tab:spatial_tracking}, such a detection moment can be as short as 1 second. In most recent prior works, the attack success is most commonly judged by first separating the entire sign detection distance range into small distance segments and claiming high attack effectiveness as long as the majority of the distance segments have high success rates~\cite{zhao2019seeing, jia2022fooling, lovisotto2021slap}. However, due to such spatial memorization, the TSR system-level hiding attack success can only be achieved when \textit{all} these distance segments have high success rates, instead of just the majority, which thus may make the TSR system-level hiding attack success harder than expected. For appearing attacks, such an impact on the system-level attack success is the \textit{opposite}, as it does not really need the majority of the segments to have high success rates; 
as long as one of them can have a high success rate, the system-level attack effect is achieved.

\begin{mtbox}{Observation\,\stepcounter{countobs}\arabic{countobs}: }
\refstepcounter{refcounter}\label{obs2}
We discover a spatial memorization design that commonly exists in today's commercial TSR systems, which can keep memorizing a sign detection result until the sign’s reaction need in the spatial domain is met (e.g., when the vehicle passes the detected sign's position). This design may create a significant discrepancy between the TSR model-level attack effect and that at the TSR system level. 
\end{mtbox}

\section{Revisiting Existing Metric and Attacks}
\label{sec:metric_revisit}

As discussed above, the newly-discovered spatial memorization design in commercial TSR systems today may create a significant discrepancy between the TSR model-level attack effect and that at the commercial TSR system level. Thus, in this section we aim to mathematically model the impact of this design on the TSR system-level attack success on both hiding and appearing attack sides, and then use the resulting new TSR system-level metrics to revisit the evaluations, designs, and capabilities of existing attacks in this problem space.

\subsection{Revisiting Existing Attack Success Metrics}
\label{sec:revisiting_existing}

\begin{figure*}[t]
\begin{minipage}[b]{0.56\linewidth}
    \footnotesize
      \centering
        \includegraphics[width=\linewidth]{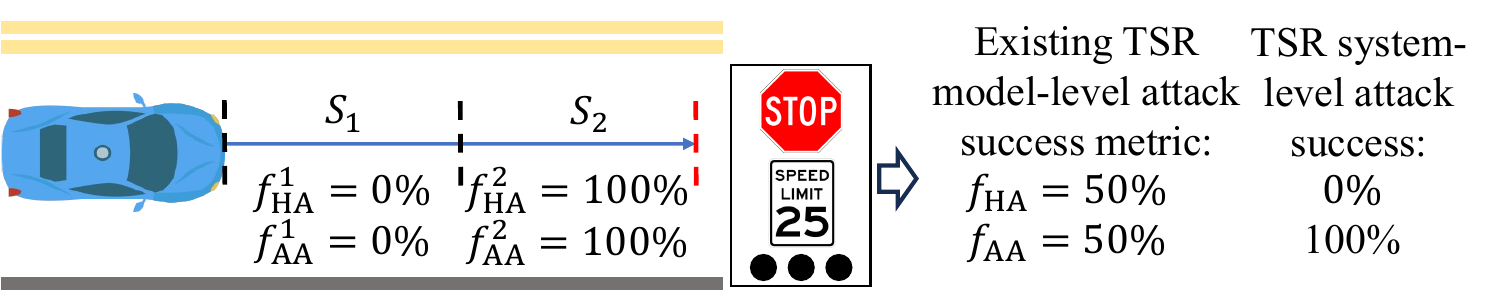} 
        \vspace{0.01in}
    \caption{Illustration of the potentially misleading effect of existing TSR model-level metrics with respect to the TSR system-level attack success. As shown, although \metricHA and \metricAA are both 50\% for this scenario, the TSR system-level attack success rates are in fact 0\% for hiding attack (HA) and 100\% for appearing attack (AA) due to spatial memorization.}
        \label{fig:limitation_prior}
        \end{minipage}
\hfill
\begin{minipage}[b]{0.42\linewidth}
\footnotesize
      \centering
        \includegraphics[width=\linewidth]{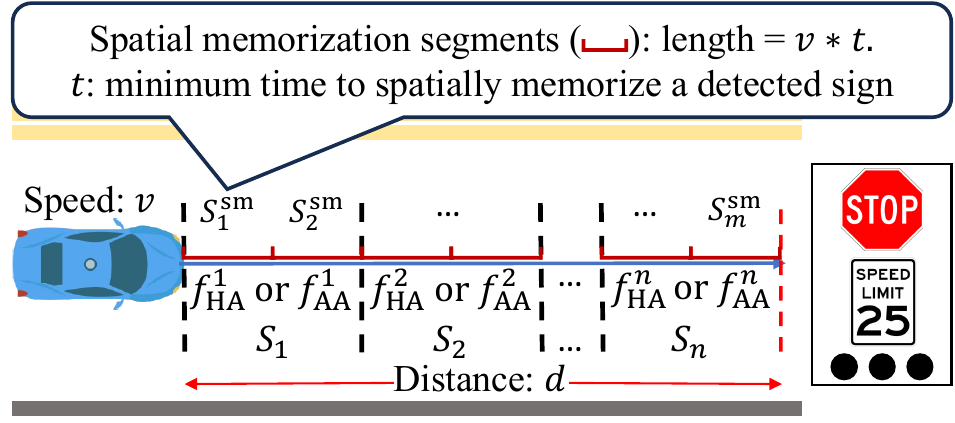}  
    \caption{Setup for calculating the proposed surrogate TSR system-level attack success metric designs (SysHA and SysAA, detailed design in~\S\ref{sec:new_metric}).}
        \label{fig:metric}
\end{minipage}
\end{figure*}

\textbf{Limitation of Existing Model-Level Attack Success Metrics.}
\label{sec:metric_limitation}
\label{sec:new_metric}
In prior works, the TSR model-level attack success metric is generally measured by averaging the per-frame attack success among a set of frames sampled in the sign detection distance range~\cite{jia2022fooling, Wang_2023_ICCV, zhao2019seeing, eykholt2018physical}. In this paper, we denote such metrics for object hiding and appearing attacks as \metricHA and \metricAA, respectively. However, such averaged per-frame attack success rates do not take the distribution of the attack effects within a targeted distance range into consideration. Due to the spatial memorization design, a certain distance range segment with an especially high or low attack success rate can directly lead to overall TSR system-level attack success or failure (as found in~\S\ref{sec:commercial_res}, the detection result can be memorized within 1 second), which can thus make such existing metrics highly misleading with respect to the TSR system-level attack effect.

Fig.~\ref{fig:limitation_prior} shows an illustrative example of such a misleading effect. As shown, there are two distance range segments, $S_1$ and $S_2$, and the attack success rates are both $f_\mathrm{HA}^1 = f_\mathrm{AA}^1 = 0\%$ for $S_1$ and $f_\mathrm{HA}^2 = f_\mathrm{AA}^2 = 100\%$ for $S_2$. As shown, using the existing \metricHA and \metricAA metrics, the attack success rates in this entire distance range will be the average of the attack success rates in $S_1$ and $S_2$, and thus are $f_\mathrm{HA} = f_\mathrm{AA} = 50\%$. However, due to spatial memorization, on the hiding attack side the 100\% $f_\mathrm{HA}^2$ cannot directly lead to the overall system-level attack success, as the TSR system can still 100\% detect the sign in $S_1$, memorize it, and correctly display it at the sign's reaction task period, leading to actually a 0\% TSR system-level attack success rate. On the appearing attack side, although $f_\mathrm{HA}^1$ is 0\%, the attack can always (100\%) trigger a fake sign detection in $S_2$ that will be memorized and displayed, and thus the end-to-end TSR system-level attack success rate is actually 100\%. As can be seen, due to spatial memorization, the TSR system-level attack success rate can be completely different from that from \metricHA and \metricAA, which can lead to highly problematic judgments of an attack's capability at the TSR system level, e.g., concluding that an attack is reasonably effective (50\%) when it actually cannot work at all (0\%).

\label{sec:metric_new_hardness}

\textbf{New Metric Design: Surrogate TSR System-Level Attack Success Metrics.}
To avoid such misleading effects of existing TSR model-level attack success metrics, the most direct solution is to perform system-level evaluations on commercial TSR systems like our efforts in~\S\ref{sec:commercial_testing}. However, it is highly difficult and also too costly for the academic community to always acquire a substantial number of commercial vehicles for experiments. And also since these commercial TSR systems are black-boxes, it is difficult to perform model design-level security research such as vulnerability cause analysis and defense evaluations. To address this, we thus propose to design \textit{surrogate} TSR system-level attack success metrics that model the spatial memorization effect on top of the existing model-level metrics, which can make them directly calculable using the more readily-accessible academic TSR model-based setups.

We start with hiding attacks. The symbols for calculating the metric are illustrated in Fig.~\ref{fig:metric}. As shown, $d$ denotes the benign-case detection distance of the targeted TSR model. In alignment with existing model-level metric calculation, we divide $d$ into $n$ \textit{measurement segments}, denoted as $S_i, i \in \{1,\ldots, n\}$, and for each segment the averaged per-frame attack success rates can be calculated, denoted as $f_\mathrm{HA}^1, f_\mathrm{HA}^2, \ldots, f_\mathrm{HA}^n$. To incorporate the spatial memorization effect, we consider the minimum time to spatially memorize a detected sign as $t$. To map this spatially memorizable detection time to the detection distance range segments, we calculate the distance traveled by the vehicle during $t$ as $v * t$, with $v$ being the vehicle speed. In this paper, we call each such distance segment of $v * t$ as \textit{spatial memorization segments}, denoted as $S^{\mathrm{sm}}_j, j \in \{1,\ldots, m\}, m = \frac{d}{vt}$. 

Due to spatial memorization, to achieve the TSR system-level attack success, a hiding attack needs to achieve attack success at every spatial memorization segment in $d$; otherwise, the sign can be detected and memorized, making the attack fail at the system level. Thus, the system-level attack success rate should be the product of \metricHA for all $S^{\mathrm{sm}}_j$, i.e., $\prod_{j=1}^{m} f_\mathrm{HA}^j$. However, to allow flexibility of the \metricHA success rate measurements, which can incur heavy-weight physical-world experiments, we do not prefer to require the measurement of \metricHA exactly at the $v * t$ granularity; instead, we aim to design the metric to be calculable for any distance choices for $S_i$. To achieve this, we thus use the $f_\mathrm{HA}^i$ to approximate the $f_\mathrm{HA}^j$ for all $S^{\mathrm{sm}}_j$ inside a measurement segment $S_i$; the resulting surrogate TSR system-level metric, which we call \textit{SysHA}, is shown in Eq.~\eqref{eq:eq2}. Note that for cases when a $S^{\mathrm{sm}}_j$ spans multiple measurement segments $S_i$, we use the average \metricHA of these segments $f_\mathrm{HA}^i$ as the \metricHA for $S^{\mathrm{sm}}_j$.
\begin{equation}
\begin{aligned}
\label{eq:eq2}
\mathrm{SysHA} = \prod_{i=1}^{n} (f_\mathrm{HA}^i) ^ \frac{m}{n} = \prod_{i=1}^{n} (f_\mathrm{HA}^i) ^ \frac{d}{n v t}
\end{aligned}
\vspace{-0.3cm}
\end{equation}

On the appearing attack side, we can use a similar design to model the impacts of spatial memorization on TSR system-level attack success. Here, the difference is that the system-level attack success can be achieved as long as the appearing attack can succeed (and thus spatially memorized) in one of the spatial memorization segments $S^{\mathrm{sm}}_j$. Thus, the probability to achieve an eventual sign appearing at the TSR system level is the negation of the probability that the appearing attack \textit{cannot} succeed (and thus memorized) in \textit{any} of the $S^{\mathrm{sm}}_j$, i.e., $1 - \prod_{j=1}^{m} (1-f_\mathrm{AA}^j)$. Following the same design in SysHA to allow measurement flexibility of \metricAA, we use $f_\mathrm{AA}^i$ to approximate $f_\mathrm{AA}^j$ for each $S^{\mathrm{sm}}_j$. The resulting surrogate TSR system-level metric for appearing attack, which we call \textit{SysAA}, is thus:
\begin{equation}
\begin{aligned}
\label{eq:eq3}
\mathrm{SysAA} = 1 - \prod_{i=1}^{n} (1 - f_\mathrm{AA}^i) ^ \frac{d}{nvt}
\end{aligned}
\vspace{-0.3cm}
\end{equation}

\textbf{Implications to TSR System-Level Attack Hardness}.
\label{sec:attack_hardness}
As informally discussed in~\S\ref{sec:commercial_res}, the spatial memorization design may make hiding attacks harder than expected and appearing attacks easier than expected. Now with the mathematical modeling of the spatial memorization's impacts on the TSR system-level attack success above, we can both theoretically and numerically analyze such impacts of the spatial memorization on the TSR system-level attack hardness.

{\bf Theorem: } When $f_\mathrm{HA}^i = f_\mathrm{AA}^i$, where $i \in \{1,\ldots, n\}$, $\mathrm{SysAA} \geqslant \mathrm{SysHA}$ always holds.

{\bf Proof.} 
To prove this theorem, Eq.~\eqref{eq:eq3} can be reformulated to the sum of the probabilities of all possible attack result scenarios that can lead to a successful system-level appearing attack. To derive that, we denote the \textit{power set} of all the measurement segments as $\mathcal{P}(S)$, where $S = \{S_i | i \in \{1,\ldots, n\}\}$. Each possible attack result scenario can be described in the form of two subsets of $S$: $A$ and $S \setminus A$, \newparts{where $A$ is a subset of $\mathcal{P}(S)$, i.e., $A \in \mathcal{P}(S)$.} Among them, the appearing attack for all $S_i \in A$ can succeed, and that for all $S_j \in (S \setminus A)$ \newparts{fails}. Due to spatial memorization, as long as $A \neq \emptyset$, the TSR system-level appearing attack effect can be achieved. Thus, SysAA can be represented as:
\begin{equation}
\begin{aligned}
\label{eq:eq4}
\mathrm{SysAA} = \sum_{A \in (\mathcal{P}(S) \setminus \emptyset)} \left( \prod_{S_i \in A} (f_\mathrm{AA}^i) ^ \frac{d}{n v t} \prod_{S_j \in (S \setminus A)} (1 - f_\mathrm{AA}^j)^ \frac{d}{n v t} \right)
\end{aligned}
\end{equation}

When $f_\mathrm{HA}^i = f_\mathrm{AA}^i$, $\mathrm{SysHA} = \prod_{i=1}^{n} (f_\mathrm{HA}^i) ^ \frac{d}{n v t} = \prod_{i=1}^{n} (f_\mathrm{AA}^i) ^ \frac{d}{n v t}$, which is actually one instance of $A \in (\mathcal{P}(S) \setminus A)$, i.e., $A = S$. 
Thus, we can calculate $\mathrm{SysAA} - \mathrm{SysHA}$:
\begin{equation}
\begin{aligned}
\label{eq:eq5}
\mathrm{SysAA} - \mathrm{SysHA} ~&= \\
\sum_{A \in \mathcal{P}(S) \setminus \{ \emptyset, S \} } & \left( \prod_{S_i \in A} (f_\mathrm{AA}^i) ^ \frac{d}{n v t} \prod_{S_j \in (S \setminus A)} (1 - f_\mathrm{AA}^j)^ \frac{d}{n v t} \right) 
\end{aligned}
\end{equation}

For $\forall A \in \mathcal{P}(S) \setminus \{ \emptyset, S \}$, $f_\mathrm{AA}^i \geqslant 0$ and $ (1-f_\mathrm{AA}^j) \geqslant 0$, where $S_i \in A$ and $S_j \in (S \setminus A)$, we can have $\prod (f_\mathrm{AA}^i) ^ \frac{d}{n v t} \prod (1 - f_\mathrm{AA}^j)^ \frac{d}{n v t} \geqslant 0 $, thus $\mathrm{SysAA} - \mathrm{SysHA} \geqslant 0$, and consequently, $\mathrm{SysAA} \geqslant \mathrm{SysHA}$. \qed

\begin{figure}[t]
    \footnotesize
      \centering
        \includegraphics[width=0.95\linewidth]{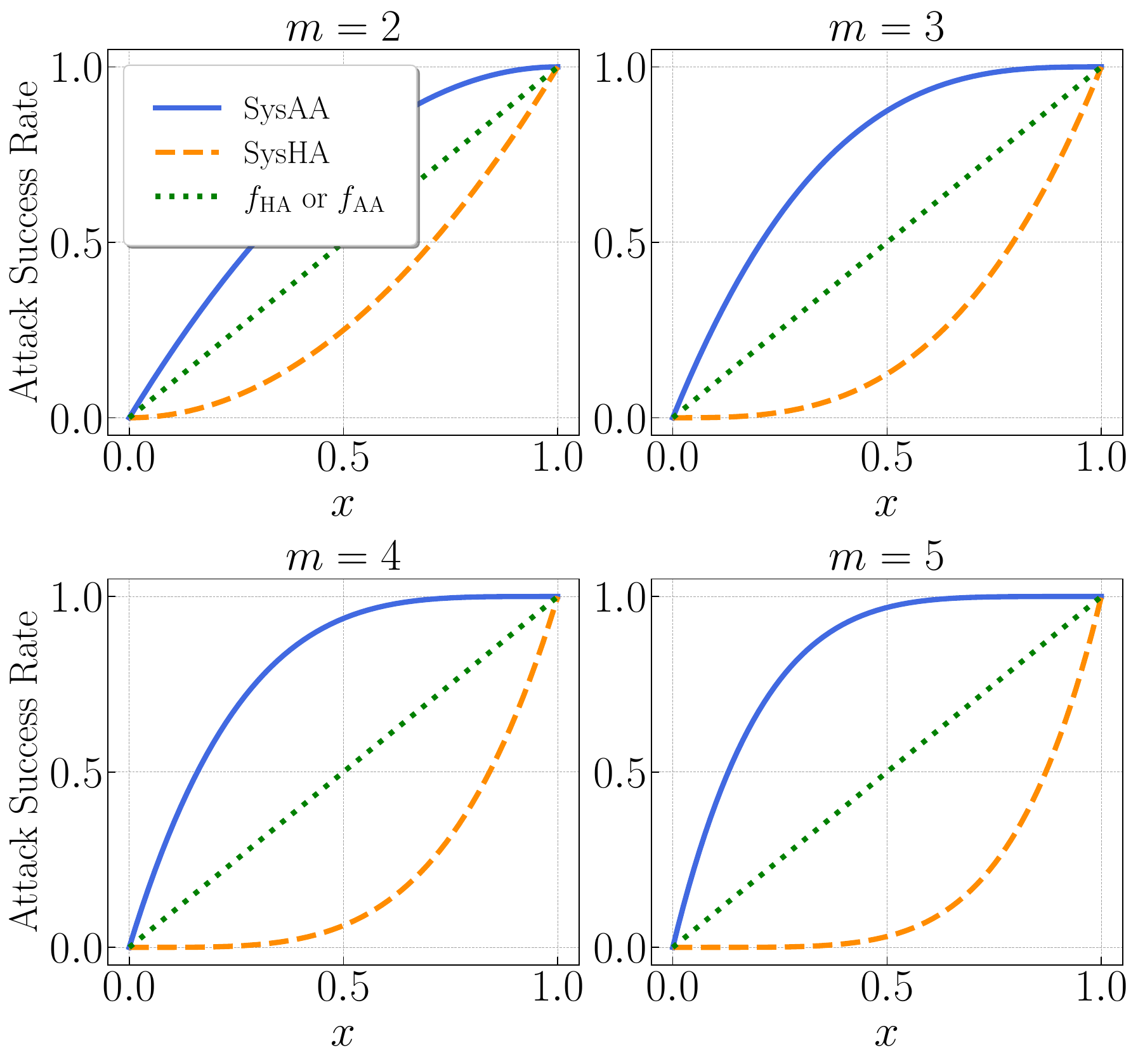} 
    \caption{Numerical analysis of the attack success rate values from SysHA, SysAA, \metricHA, and \metricAA when $m = 2, \ldots, 5$. Here, we set $n=m$ and $x = f_\mathrm{HA}^i = f_\mathrm{AA}^i$ for all $i \in \{1,\ldots, n\}$.}
        \label{fig:experimental}
\end{figure}

{\bf Numerical Analysis.} The theoretical analysis above can mathematically reveal that SysAA can always be equal or larger than SysHA when \metricHA and \metricAA are equal. However, it cannot reveal how large the gap between SysHA and SysAA can be. Thus, we further perform a numerical analysis of SysHA and SysAA. In this analysis, we set $n = m$, and $x = f_\mathrm{HA}^i = f_\mathrm{AA}^i$ for all $i \in \{1,\ldots, n\}$. Thus, $\mathrm{SysHA} = x^m$ and $\mathrm{SysAA} = 1- (1-x)^m$. In Fig.~\ref{fig:experimental}, we plot the values of SysHA, SysAA, \metricHA, and \metricAA when $m = 2, \ldots, 5$. This is a realistic approximation of $m$ since as found in~\S\ref{sec:commercial_res}, commercial TSR systems can spatially memorize a detection result within one second. At normal driving speed, the traffic sign can appear for at least 2-5 seconds, and thus there at least exist 2-5 spatial memorization segments. As shown, SysAA is always greater than SysHA when $x$ is between 0 and 1, and the difference ($\mathrm{SysAA} - \mathrm{SysHA}$) is at least 50\% (when $m=2$) and can be as high as 93.8\% when $m=5$. This means that even when hiding attacks and appearing attacks seem to have similar model-level attack effectiveness, due to the spatial memorization design the TSR system-level attack effectiveness can have huge differences ($\geqslant$93.8\% differences in absolute attack success rate values). 

Meanwhile, we can also observe that SysHA and SysAA can both differ significantly from \metricHA and \metricAA results. As shown, when $m=5$, SysAA can be much higher than \metricAA (46.9\% difference in absolute attack success rate values), and SysHA can be much lower than \metricHA (also 46.9\% difference in absolute attack success rate values). This thus numerically proves the potentially misleading effect of existing TSR model-level metrics with respect to the TSR system-level attack effect, which can lead to a misjudgment of the attack effectiveness to the extent of nearly 50\% in absolute attack success rate values.

\newparts{\textbf{About the novelty and importance of the new metric design.} Note that the metric and the concept of spatial memorization might seem straightforward, but no prior research has discovered and analyzed the impact of this important TSR system-level design from the security perspective, nor formulated it mathematically. Before this paper, it was unknown that such a design is commonly used in commercial TSR systems. Additionally, as discussed above, mathematically modelling and quantifying its impacts on adversarial attack success rates as a metric is crucial for the research community to systematically understand real-world system vulnerabilities, especially from the commercial TSR systems perspective. Meanwhile, our theoretical and empirical analyses above further provide mathematically provable insights about the design-level implications and the magnitude of such impacts from the spatial memorization design. These thus all make our newly-proposed surrogate TSR system-level attack success metrics valid and significant scientific contributions.

Meanwhile, we would also like to clarify that we do not intend to claim that the existing TSR model-level metrics do not have important value when compared with our proposed surrogate TSR system-level metric. As explained above, the design of this new metric and the later experimental comparisons of the results from these two metrics are only for the purpose of scientifically understanding how much the newly-observed spatial memorization design in commercial TSR systems today can affect the judgment and understanding of the capability of a certain attack design at the TSR system level.}

\begin{mtbox}{Observation\,\stepcounter{countobs}\arabic{countobs}: }
\refstepcounter{refcounter}\label{obs3}
Due to spatial memorization, hiding attacks are theoretically harder (if not equally hard) than appearing attacks in achieving TSR system-level attack success. Such an attack hardness gap can be huge (e.g., $\geqslant$93.8\% absolute differences in the attack success rate values). Meanwhile, due to the lack of consideration of spatial memorization, existing TSR model-level attack success metrics can be highly misleading in judging the TSR system-level attack effect, with a potential of having $\sim$50\% absolute attack success rate value differences.
\end{mtbox}

\subsection{Revisiting Existing Attacks}
\label{sec:existing_attack_HA}
\label{sec:existing_attack_AA}

In this section, We use the SysHA and SysAA metrics to revisit the evaluations, designs, and attack capabilities of the important prior works in this problem space.

Specifically, we revisit (1) {both hiding and appearing attacks}, with RP$_2$~\cite{eykholt2018physical}, SIB~\cite{zhao2019seeing}, FTE~\cite{jia2022fooling} as representative examples on the hiding side, and SIB~\cite{zhao2019seeing} and DM~\cite{sato2023intriguing} as representative examples on the appearing side; (2) {both white-box and black-box transfer attack setups}, with the original white-box attack setups evaluated in the original papers (e.g., YOLO v2 (Y2) for RP$_2$, YOLO v3 (Y3) for SIB, and YOLO v5 (Y5) for FTE) on the white-box side, and a set of representative transfer target models on the black-box transfer attack setup side. Specifically, this transfer target model set includes 6 models in total, 3 from the one-stage model design family (YOLO v8 (Y8)~\cite{yolov8}, YOLOS (YS)~\cite{fang2021you}, DETR~\cite{carion2020end}) and 3 from the two-stage model design family (two versions of Faster RCNN with different backbones~\cite{torchmodel}, and Mask RCNN~\cite{torchmodel}).

\newparts{In all the experiments, we focus on STOP sign as the representative attack target since it is used the most commonly in prior works in their physical-world evaluation (in Table~\ref{tab:survey}, five out of the six prior works only used STOP sign in their physical-world experiments for TSR), which can thus best suit our need in this section to revisit the evaluation of prior works.} All the attacks are physically printed out and measured outdoor. The distance range measured is 0 to 30m following the setups in prior works~\cite{eykholt2018physical, zhao2019seeing, jia2022fooling}. For each attack setup, we calculate the SysHA or SysAA attack success rate by averaging the SysHA or SysAA values across the full combinations of a set of common speeds (25 mph, 30 mph, and 35 mph, the most common speed limits for STOP sign-controlled roads~\cite{Wang_2023_ICCV}) for $v$ and all possible minimum spatial memorization time $t$ (0.05 seconds to 1 second with a step of 0.05 seconds considering the common camera frame rate 20Hz~\cite{apollo} and the 1-second upper-bound of $t$ found from our experiments in~\S\ref{sec:commercial_res}). 

\begin{table}[t]
\tabcolsep 0.03in
\small
    \caption{White-box attack effectiveness for representative prior works on hiding attacks measured by \metricHA and SysHA. As shown, while the \metricHA results can indeed be claimed as effective, it may not be appropriate to claim so with SysHA. This suggests a potential lack of TSR system-level attack effectiveness even in the white-box setting for existing works.}
    \centering
    \begin{tabular}{ccccccccc}

    \toprule
    & \multicolumn{7}{c}{\metricHA} &  \\
    \cmidrule(lr){2-8}
    & \multicolumn{6}{c}{Distance ranges (meters)} &  & \\ \cmidrule(lr){2-7}
    & 0-5 & 5-10 & 10-15 & 15-20 & 20-25 & 25-30 & \multirow{-2.5}{*}{Ave.}& \multirow{-4}{*}{SysHA}\\
    \midrule
    RP$_2$ & 41.8\% & 10.0\% & 23.8\% & 65.4\% & 99.9\% & 100\% & 56.8\% & 6.6\%\\
    \midrule
    SIB & 84.6\% & 56.6\% & 82.0\% & 99.2\% & 100\% & 100\% & 87.1\% & 45.1\%\\
    \midrule
    FTE  & 88.9\% & 57.1\% & 13.6\% & 3.1\% & 47.8\% & 74.5\% & 47.5\% & 5.2\%\\
    \bottomrule
         
    \end{tabular}
    \label{tab:table5}
\end{table}

In the following, we report the most notable findings our these revisiting experiments. 

\textbf{White-Box Attack Effectiveness.} We start by revisiting the prior works in white-box attack setups. The impact of the spatial memorization design on the TSR system-level attack success is orthogonal to the attacker's knowledge level (e.g., white- or black-box) of the targeted TSR model, and thus it is of interest to use our new metrics to revisit existing attacks even in the white-box attack setting. Table~\ref{tab:table5} shows the white-box attack effectiveness of representative existing hiding attacks measured by both \metricHA and SysHA. When using \metricHA, the attack success rates are similar to those reported in the original papers~\cite{eykholt2018physical, zhao2019seeing, jia2022fooling} and can be claimed as effective (from $\sim$50\% to 90\%). However, as numerically analyzed in~\S\ref{sec:new_metric}, due to spatial memorization the TSR system-level attack success rate can be much lower. As shown, when using SysHA, the attack success rates decrease significantly by 2 to 9 times, with the absolute success rate value drops of at least 40\%. Most notably, the attack success rates for RP$_2$ and \newparts{FTE} are dropped to at most 6.6\%. Thus, it may not be appropriate to claim that these attacks can be effective at the TSR system level. This suggests a potential lack of TSR system-level attack effectiveness for existing works even in the white-box settings. 

\begin{mtbox}{Observation\,\stepcounter{countobs}\arabic{countobs}: }
\refstepcounter{refcounter}\label{obs4}
When spatial memorization is considered, prior works in this problem space may not be claimed as effective at the TSR system level even in white-box attack settings. Our newly-proposed surrogate TSR system-level metrics can help improve this in the future as they can be leveraged to better approximate the impact of spatial memorization on the TSR system-level attack success.
\end{mtbox}

\begin{table*}[t]
\small
    \caption{Black-box transfer attack effectiveness for representative prior works on hiding attacks measured by \metricHA and SysHA. \newparts{Each success rate value is an average over the results from 6 representative transfer target models: 3 one-stage models (Y8, YS, DETR) and 3 two-stage models (two Faster RCNN with different backbones, MaskRCNN), which are detailed more in~\S\ref{sec:existing_attack_HA}.} As shown, although the \metricHA values show descent transfer attack success rates similar to those reported by the original paper, those calculated by SysHA are much lower, similar to the observations from the commercial TSR testing in~\S\ref{sec:commercial_res} and Table~\ref{tab:results_ss}.}
    \centering
    \begin{tabular}{cccccccccc}

    \toprule
    & & \multicolumn{8}{c}{Transfer attack success rates (averaged over a set of six transfer target models (\S\ref{sec:existing_attack_HA})}\\
    \cmidrule(lr){3-10}
    & & \multicolumn{7}{c}{\metricHA} &  \\
    \cmidrule(lr){3-9}
    & \multirow{-2}{*}{\shortstack{Original paper \\ transferability}}& 0-5m & 5-10m & 10-15m & 15-20m & 20-25m & 25-30m & Ave. & \multirow{-2.5}{*}{SysHA}\\
    \midrule
    RP$_2$ & 18.9\% & 36.4\% & 32.0\% & 29.6\% & 46.0\% & 61.3\% & 50.0\% & 42.6\% & 14.5\%\\
    \midrule
    SIB & 46.1\% & 20.7\% & 26.5\% & 37.2\% & 42.6\% & 54.9\% & 51.2\% & 38.9\% & 12.4\%\\
    \midrule
    FTE & 89.8\% & 29.2\% & 36.4\% & 29.3\% & 34.0\% & 45.5\% & 40.1\% & 35.7\% & 11.0\%\\
    \midrule
    \midrule
    Ave. & 51.6\% & 28.8\% & 31.6\% & 32.0\% & 40.9\% & 53.9\% & 47.1\% & 39.1\% & 12.6\% \\
    \bottomrule
         
    \end{tabular}
    \label{tab:table6-2}
\end{table*}

\textbf{Black-Box Transfer Attack Effectiveness.} We next revisit the black-box transfer attack effectiveness of existing works. Table~\ref{tab:table6-2} shows the results for representative prior works on hiding attacks measured by both \metricHA and SysHA. As shown, when using \metricHA, all prior works show a descent transfer attack success rates at $\sim$40\%, which is very similar to the reported numbers from the original papers on average ($\sim$50\%). However, when using SysHA, the success rates becomes much lower, which are generally decreased by around 3 times for all three attacks. This is in fact quite similar to the observations from the commercial TSR testing in~\S\ref{sec:commercial_res} and Table~\ref{tab:results_ss}, which found that the black-box transfer attack success rates against commercial TSR systems are almost a magnitude lower than those reported by the original papers on average. This suggests that spatial memorization may indeed be a major factor for the observed much lower-than-expected black-box transfer attack success rates against commercial TSR systems.

\begin{mtbox}{Observation\,\stepcounter{countobs}\arabic{countobs}: }
\refstepcounter{refcounter}\label{obs5}
\newparts{When spatial memorization is considered, the black-box transfer attack success rates of prior works at TSR system level can be much lower than expected (only $\sim$13\%) for hiding attack. This suggests that existing hiding attack works are unlikely to have direct impacts on real-world commercial TSR systems in general, which is consistent with our observations in our large-scale commercial TSR systems testing. This suggests that future work in this problem space should focus more on black-box attack settings, which can be more generally enabled by our newly-proposed surrogate TSR system-level attack success metrics.}
\end{mtbox}

\begin{table}[t]
\tabcolsep 0.06in
\small
    \caption{Attack capability comparison between the hiding and appearing attacks proposed from the same prior work (SIB~\cite{zhao2019seeing}) measured by \metricHA, SysHA, \metricAA, and SysAA. As shown, if using the prior TSR model-level attack success metrics (\metricHA, \metricAA), the judgment of the attack capability differences between the proposed hiding and appearing attacks can be the \textit{completely opposite} to those using SysHA and SysAA due to the consideration of spatial memorization.}
    \centering
    \begin{tabular}{ccccc}

    \toprule
    & \multicolumn{2}{c}{Hiding attack} & \multicolumn{2}{c}{Appearing attack} \\
    \cmidrule(lr){2-3}
    \cmidrule(lr){4-5}
    SIB~\cite{zhao2019seeing} & \metricHA & SysHA & \metricAA & SysAA \\
    \midrule
    White-box attack & 87.1\% & 45.1\% &  29.1\% & 87.6\%\\
    Black-box transfer attacks & 38.9\% & 12.4\% & 31.7\% & 64.2\%\\
    \bottomrule
         
    \end{tabular}
    \label{tab:table8-1}
\end{table}

\textbf{Hiding vs. Appearing Attack Effectiveness.} As numerically analyzed in~\S\ref{sec:new_metric}, it can be much harder to achieve hiding attack success than to achieve appearing attack success at the TSR system level due to spatial memorization. Now with concrete attack examples from prior works, we can more directly study such attack hardness gaps. Table~\ref{tab:table8-1} shows the comparison of the attack success rates for the hiding and appearing attacks from the same prior work SIB~\cite{zhao2019seeing}. The success rates are calculated for both prior TSR model-level metrics (\metricHA, \metricAA) and our newly-proposed surrogate TSR system-level metrics (SysHA, SysAA) under both white-box and black-box transfer attack settings. Here, the black-box transfer attack success rates are calculated by averaging results over a set of 6 representative target models (\S\ref{sec:existing_attack_HA}).

As shown, in both white-box and black-box transfer attack settings, SysHA is lower than \metricHA and SysAA is higher than \metricAA, which is consistent with our numerical analysis results (\S\ref{sec:new_metric}) and is caused by the spatial memorization effect. Interestingly, if we only use prior TSR model-level metrics (\metricHA, \metricAA) to judge the attack capabilities between the proposed hiding and appearing attacks, the conclusion will be that the proposed hiding attack is more effective (if not much more effective) than the proposed appearing one, as the \metricHA results are always higher than the \metricAA results in both white-box and black-box transfer attack settings; in particular, in the white-box setting, the \metricHA success rate is $\sim$3 times of the \metricAA one. However, when using SysHA and SysAA, the conclusion is the \textit{completely opposite}: the appearing attack success rate is always higher than the hiding attack one in both white-box and black-box transfer attack settings, and the former can be $\sim$2-5 times higher. This suggests that if not considering spatial memorization, the judgment of the TSR system-level attack capabilities across hiding and appearing attacks can be completely wrong. This further highlights the importance of systematically modelling the impacts of spatial memorization on the TSR system-level attack success, which is exactly what we aim to achieve in the design of SysHA and SysAA.

\begin{mtbox}{Observation\,\stepcounter{countobs}\arabic{countobs}: }
\refstepcounter{refcounter}\label{obs6}
Using the hiding and appearing attacks proposed from the same prior work, the hiding one can indeed be much harder (2-5 times) than the appearing one in both white-box and black-box transfer attack settings after spatial memorization is considered. However, if using the prior TSR model-level attack success metrics, the judgment of such relative attack hardness differences can be the completely opposite, which thus highlights the necessity of the design of the new SysHA and SysAA metrics.
\end{mtbox}

\begin{table}[t]
\tabcolsep 0.03in
\small
    \caption{The white-box appearing attack effectiveness of RP$_2$~\cite{eykholt2018physical} with and without the Nested AE (NAE) design measured by \metricAA and SysAA. As shown, when using \metricAA, the NAE design can indeed ``significantly improve the robustness of adversarial attack in various distances'' according to the original paper that proposed NAE as the key new design contribution~\cite{zhao2019seeing}: it improves \metricAA in every distance range segments, leading to a 22\% improvements on average). However, when using SysAA with spatial memorization considered, the success rate improvement is almost negligible ($\sim$1\%).}
    \centering
    \begin{tabular}{cccccccc}
    \toprule
     & \multicolumn{6}{c}{\metricAA} &  \\
    \cmidrule(lr){2-7}
    \multirow{-2.5}{*}{RP$_2$~\cite{eykholt2018physical}} & 0-5m & 5-10m & 10-15m & 15-20m & 20-25m & Ave. & \multirow{-2.5}{*}{SysAA}\\
    \midrule
   w/o NAE & 86.8\% & 100\% & 64.7\% & 66.9\% & 19.5\% & 67.6\% & 98.2\% \\
   w/ NAE & 100\% & 100\% & 100\% & 88.3\% & 25.8\% & 82.8\% & 100\%\\ 
    \bottomrule   
    \end{tabular}
    \label{tab:table10-1}
\end{table}

\textbf{Judgement of the Value of New Attack Designs.} As numerically analyzed in~\S\ref{sec:new_metric}, spatial memorization can significantly impact the attack success at the TSR system level, with a potential of having $\sim$50\% absolute attack success rate value difference. Due to this, it is possible that certain new attack designs proposed in prior works can be seemingly highly beneficial to the attack success when judged using the prior TSR model-level metrics, but when judged with the consideration of spatial memorization, the benefits to the attack success are actually very minimal. This thus may significantly change how we judge the practical value of certain prior attack designs at the TSR system level. Table~\ref{tab:table10-1} shows our investigation into one such example. As shown, in this experiment we compare the white-box appearing attack effectiveness of RP$_2$~\cite{eykholt2018physical} with and without a specific attack design called Nested AE, which we denote as \textit{NAE}. This design is proposed by Zhao et al.~\cite{zhao2019seeing} as a key new appearing attack design contribution. In this design, the key idea is to decouple the task of varying distance attack into two pieces: long-distance and short-distance attacks, and distribute them to different regions on the adversarial pattern printed on a whote-sign poster, with the goal of systematically increasing the distance of an appearing attack~\cite{zhao2019seeing}. 

\begin{figure}[t]
    \footnotesize
    \vspace{0.2in}
      \centering
        \includegraphics[width=0.7\linewidth]{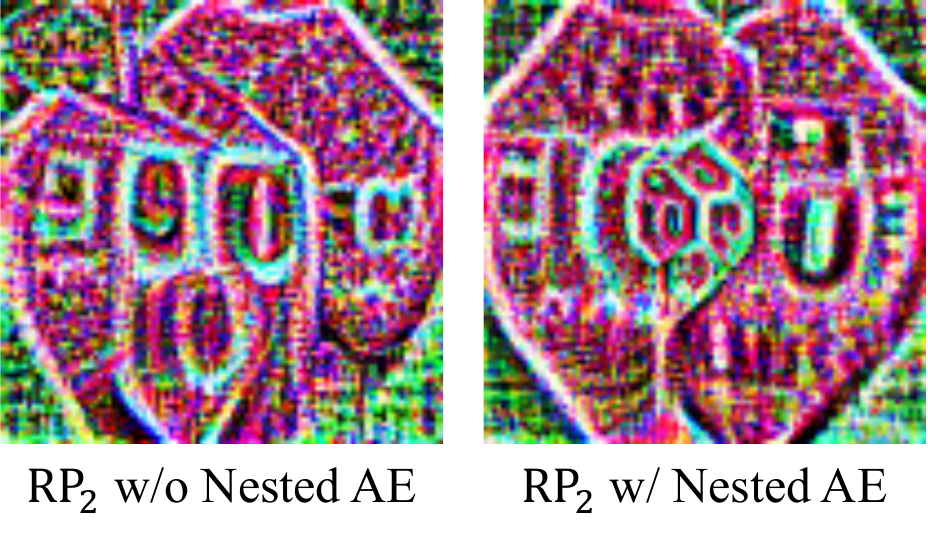}  
    \caption{Visualisation of the adversarial patterns with and without the Nested AE (NAE) design reproduced by us using RP$_2$ (the original design has not been open-sourced). As shown, the ``nested'' adversarial pattern feature is highly consistent with that reported in the original paper (e.g., Fig.~5 in~\cite{zhao2019seeing}).}
        \label{fig:visual_nestae}
\end{figure}

Fig.~\ref{fig:visual_nestae} shows a visualization of the adversarial patterns with and without the NAE design reproduced by us (like what we reported in~\S\ref{sec:setup} for prior works on appearing attacks, this prior attack design also has not been open-sourced so far). As shown, we are able to reproduce the designed adversarial pattern features, with an inner STOP sign-like adversarial pattern ``nested'' at the center of an outer adversarial pattern, with the inner one for short-distance attack effect and the outer one for long-distance attack effect, which is consistent with the reported adversarial pattern features in the original paper (can be seen by Figure 5 in~\cite{zhao2019seeing}). As shown by our results in Table~\ref{tab:table10-1}, when using \metricAA, such a design can indeed ``significantly improve the robustness of adversarial attack in various distances'' as claimed in the original paper~\cite{zhao2019seeing}: it improves \metricAA in every distance range segments from 0 to 30m, which is able to increase the average \metricAA success rates by 22\%, from lower than 70\% to higher than 80\%. However, when using SysAA with spatial memorization considered, the success rate improvement is \textit{almost negligible} \newparts{($\sim$2\%), from 98.2\% to 100\%. This is mainly because due to spatial memorization, it can be much easier to achieve a TSR system-level appearing attack success compared to the attack capabilities that can be reflected by the direct TSR model-level metrics like \metricAA. Thus, even with a naive attack design, the attack success rates at the TSR system level can already be very high, which can thus make the necessity and practical value of more sophisticated appearing attack designs very low.}

\begin{mtbox}{Observation\,\stepcounter{countobs}\arabic{countobs}: }
\refstepcounter{refcounter}\label{obs7}
Due to spatial memorization, the benefits of certain attack designs can be seemingly high (e.g., $>$20\% attack success rate increase) using prior TSR model-level success metrics, but actually nearly negligible (e.g., only 1\% increase) at the TSR system level. This thus highlights the necessity of attack success metrics that can incorporate the impact of spatial memorization, such as SysHA and SysAA, with regard to the judgment of the necessity and practical value of any given attack designs in this problem space, for both the designs in the past or those in the future.
\end{mtbox}

\section{Discussions}
\label{sec:discussions}
In this section, we provide detailed discussion on the ethics, limitations, and future work.

\subsection{Ethics}
\label{sec:ethics}
In discussing the ethical considerations of our measurement study, it is crucial to highlight the measures taken to ensure safety and responsibility. Our experiments with commercial vehicles are conducted on the roof of a parking structure, a controlled environment where we can ensure the absolute absence of other vehicles and people when performing the experiments. 
Additionally, we take precautions to make sure that the adversarial attacks are not visible anywhere from public roads, thereby eliminating any risk to others.

Recognizing the direct impacts of our study on the security of commercial vehicles, we have performed responsible vulnerability disclosure to all the potentially affected vehicle manufacturers following the ethical standard in the security community. Specifically, this involved informing the vehicle manufacturers for all the vehicle models we tested (i.e., C1 to C4) about our testing results, especially those with vehicle models that were tested vulnerable such as the the vehicle manufacturer for C2; we have already done all these.
This proactive approach allows the manufacturers to address and mitigate any potential risks posed by these attacks, ensuring the safety and security of their vehicles. 
Additionally, even in this submission version, we do not directly reveal the exact vehicle brands and models for C1 to C4 to protect the potentially-affected companies  (\S\ref{sec:commercial_testing}).

\subsection{Limitations and Future Work}
\label{sec:limitation}
\newparts{
\textbf{Threats to Validity for the Statistical Significance in Commercial Systems Testing Results.} To scientifically interpret our commercial system testing results, it is important to understand their statistical significance. First, for the overall transfer attack success rates against commercial systems over all 30 attack test combinations, the result (6.67\% in Table~\ref{tab:results_ss}) is statistically significant with $p<0.02$ using Z-Test~\cite{lawley1938generalization}, Binomial Test~\cite{wagner2005binomial}, One-Sample T-Test~\cite{ross2017one}, and Wilcoxon Signed-Rank Test~\cite{woolson2005wilcoxon}. Second, for the result of each test combination, we technically cannot compute the statistical significance values since the variance for each is 0 (all failure or all success as shown in Table~\ref{tab:results_ss}) and statistical testing methods are designed for data samples with variance (and thus their calculation generally requires division by the standard deviation~\cite{2020SciPy-NMeth, wagner2005binomial, lawley1938generalization, woolson2005wilcoxon, ross2017one, st1989analysis, schumacker2013f, macfarland2016kruskal, hsu2014paired, bollen1981pearson}, which is 0 in our case). For our case, this may not be addressable by simply increasing the number of attempts, since the root cause is the lack of variation in the results, which is likely to continue since the output of each test run for us is binary (failure or success). In hypothesis testing, if all observed values are exactly the same as the hypothesized value (i.e., there is no variation in the data), it means that the data perfectly matches the null hypothesis~\cite{shaver1993statistical} and there is no need to conduct further statistical tests to determine significance, as the data inherently confirms the hypothesis~\cite{frost2020hypothesis, nullHyp}. In our experiments, this is indeed the case as (1) we observe no variations during the 3 attempts and (2) considering the spatial memorization design, in each attempt there are also multiple TSR model-level tests when the vehicle approaches the sign, but still no variations were observed, which made us believe that the results are unlikely to change even if we try a few more times (we cannot afford a significant increase in the number of attempts as we explain below). Note that this is also already statistically more rigorous than all prior works in this: they only tried once for each attack test~\cite{sato2023intriguing, jia2022fooling} (we have confirmed this with the authors), which is not enough to even calculate variance.

Nevertheless, it is indeed possible that if we significantly increase the number of attempts, some variations will appear and thus allow us to calculate statistical significance. However, since we need to manually drive the car past the sign and circle back to the same starting point for each attempt, and also need to manually take down and put up new adversarial patterns, at 3 attempts and 14 different tested signs (2 benign, 12 attacks) we need to spend over 2 hours per vehicle model. If we largely increase the attempt number, it may not be possible to control the lighting conditions to be comparable across these tests, which is one of the most critical factors for the effectiveness of physical-world adversarial attacks~\cite{zhao2019seeing, jia2022fooling, xue2021naturalae}. For example, if we conduct at least 30 attempts per test as commonly suggested in statistical testing~\cite{thirty-time}, it will cost at least 10 hours in total and the lighting conditions will be completely changed. Note that this is an inherent limitation for any outdoor experimental setup on commercial vehicles. To address this, one possibility is to use an indoor vehicle testing facility with controlled lighting conditions. However, we are not aware of any prior works in AD security that can have such a setup, and also it is unclear whether such a setup can accurately simulate outdoor conditions, as the targeted physical environment for TSR adversarial attacks is outdoor. We thus leave the solution exploration of this to future work.}

\textbf{More Root Cause Analysis for Commercial TSR System Testing Results.} In this work, we were able to perform the first large-scale measurement of existing physical-world adversarial attacks on commercial TSR systems by performing black-box transfer attack testing on the commercial systems. Although this was able to fill the critical research gap in achieving a more general understanding of existing attacks' impacts on real-world commercial TSR systems, certain aspects remain partially understood 
\newparts{, for example the reason why the TSR function of certain commercial systems can actually be much more vulnerable than academic TSR models, and also why the two successful attack setups are limited to the STOP sign detection. There are multiple possible causes we can speculate, e.g., for the former due to the need to reduce false alarm rates in real-world deployment, or due to the long deployment cycle to integrate latest academic model designs; and for the latter due to the need to customize the detection accuracy for certain sign types because of real-world deployment or customer needs. However, due to the black-box nature of these commercial systems, it requires more systematic follow-up studies to scientifically answer these questions, which we believe is one of the most highly desired future works.
}

\textbf{Defense-Side Explorations.} In this work, we mainly focus on measuring and understanding the potential gap between existing academic research and real-world commercial systems on the attack side. In the future, it is also of interest to explore such a potential gap on the defense side. Such a gap may indeed exist, for example, due to spatial memorization, hiding attacks can be easier to be defended against at the TSR system level than expected, since as long as the defense method can prevent the hiding attack success in \textit{any} of the spatially memorizable sign detection periods, it can prevent the hiding attack success at the TSR system level. For appearing attacks, this will be the opposite since to prevent the appearing attack success at the TSR system level, the defense method has to prevent the appearing attack success in \textit{all} of the spatially memorizable sign detection periods. We thus leave a systematic investigation of these aspects to the future work.

\newparts{
\textbf{About misclassification attacks.} In this work, we focus on the two most representative TSR adversarial attack types so far with highly demonstrated physical-world realism, hiding and appearing attacks (\S\ref{sec:TSR_attack}). Meanwhile, there are also prior works studying misclassification attacks on TSR models (i.e., changing the detection from one sign to another)~\cite{jia2022fooling, eykholt2018robust}, which is less studied in the security community (potentially due to their relatively indirect attack consequences compared to direct sign hiding or appearing) but can also be of interest to be studied from the commercial systems perspective, especially if we consider the potential impact from spatial memorization. Specifically, for such attacks, if the attack doesn’t always succeed/fail in every spatial-memorization segment, the detected sign class will change across spatial-memorization segments. The impact of spatial memorization will depend on whether a later-detected different sign class will override a previously-memorized sign class. For example, if the design is to not override, the TSR system-level success depends on the model-level success of the first/farthest-to-the-sign spatial-memorization segment. If the design is to override, for speed limit signs the TSR system-level success depends on the last/nearest-to-the-sign spatial-memorization segment, while for STOP sign the driver will see alternating/simultaneous display of correct and incorrect signs and thus it may require to newly define the TSR system-level success from the driver’s perspective. To effectively answer these new questions, a separate follow-up study is required since it will require new research methodology designs starting from the commercial system testing stage, which we thus leave to future work. 
}

\newparts{
\textbf{Impact from Other Possible Data Sources for TSR.}
Our current commercial TSR system testing results are unlikely to be influenced by other possible sources for TSR such as GPS/map. For example, since all the tested signs are temporarily placed at rooftop of a parking structure, no existing maps can have these sign information. Thus, the failed attack attempts cannot be due to possible sign information retrieved from GPS/map. Meanwhile, out of the 13 car brands providing TSR in Table 1, for 12 of them (including all the brands we tested) we cannot find any evidence that fusion is used; specifically, for them their official websites or vehicle manuals only mention the use of camera without any mention of the use of other sources. Nevertheless, systematically understanding whether fusion can be an impact factor on the current results (and if so, how much) can be an interesting follow-up research direction, especially those on designing new analysis methods to systematically understand the root causes of many current observations given the black-box nature of these systems as discussed above. We thus leave this to future work.}

\label{sec:futurework}

\section{Conclusion}
\label{sec:conclusion}

In this paper, we conduct the first large-scale measurement of physical-world adversarial attacks against commercial TSR systems. Our testing results reveal that although it is possible for existing attack works from academia to have highly reliable (100\%) attack success against certain commercial TSR system functionality, such black-box commercial system attack capabilities are not generalizable, leading to a much lower-than-expected black-box transfer attack success rates overall. We find that one potential major factor is the spatial memorization design that commonly exists in today's commercial TSR systems. We design new attack success metrics that can mathematically model the impacts of this design on the TSR system-level attack success, and use them to revisit existing attacks. Through these efforts, we uncover 7 novel observations, some of which can directly challenge the observations or claims in prior works due to the use of our new metrics. 
We hope that the results and new insights from this work can help inspire and facilitate more practically meaningful and impactful research in this critical problem space.

\section*{Acknowledgments}
\label{sec:ack}
We would like to thank Ziwen Wan, Tong Wu, Junze Liu, Fayzah Alshammari, Justin Yue, and the anonymous reviewers for their valuable and insightful feedback. This research was supported by the NSF under grants CNS-1929771 and CNS-2145493; USDOT under Grant 69A3552348327 for the CARMEN+ University Transportation Center.

\bibliographystyle{IEEEtran}
\bibliography{main}

\end{document}